\documentclass[10pt,twocolumn,letterpaper]{article}

\usepackage[pagenumbers]{cvpr} 

\usepackage{graphicx}
\usepackage{amsmath}
\usepackage{amssymb}
\usepackage{booktabs}
\usepackage{url} 
\usepackage{float}
\usepackage{color}
\usepackage{bm}
\usepackage{bbm}
\usepackage{algorithm}
\usepackage{algpseudocode}
\usepackage{enumitem}
\usepackage{siunitx}
\usepackage{multirow}
\usepackage{amsfonts}       
\usepackage{nicefrac}       
\usepackage{microtype}      
\usepackage{xcolor}         
\usepackage{diagbox} 
\usepackage{tabularx}
\usepackage{hhline}
\usepackage{arydshln}		
\usepackage{mathtools} 
\usepackage{amsthm, soul}			

\def\defn{\,\coloneqq\,}
\def\argmax{\mathop{\mathrm{arg\,max}}} 
\def\argmin{\mathop{\mathrm{arg\,min}}} 
\def\prox{\mathrm{\bf Prox}} 

\def\cbm{{\bm{c}}}

\def\xbm{{\bm{x}}}

\def\ybm{{\bm{y}}}
\def\zbm{{\bm{z}}}

\def\thetabm{{\bm{\theta}}}
\def\zerobm{\bm{0}}

\def\Abm{{\bm{A}}}

\def\epsilonbm{{\bm{\epsilon}}}
\def\Ibf{{\mathbf{A}}}
\def\Ibf{{\mathbf{B}}}
\def\Ibf{{\mathbf{C}}}
\def\Ibf{{\mathbf{D}}}
\def\Ibf{{\mathbf{E}}}
\def\Ibf{{\mathbf{F}}}
\def\Ibf{{\mathbf{G}}}
\def\Ibf{{\mathbf{H}}}
\def\Ibf{{\mathbf{I}}}

\def\xbmtilde{{\widetilde{\bm{x}}}}

\def\R{\mathbb{R}}
\def\E{\mathbb{E}}

\def\Ncal{{\mathcal{N}}}

\def\Fcal{{\mathcal{F}}}

\definecolor{lightgreen}{rgb}{.9,1,.9}
\definecolor{lightblue}{rgb}{.9,.9,1.}

\usepackage[pagebackref,breaklinks,colorlinks]{hyperref}

\usepackage[capitalize]{cleveref}
\crefname{section}{Sec.}{Secs.}
\Crefname{section}{Section}{Sections}
\Crefname{table}{Table}{Tables}
\crefname{table}{Tab.}{Tabs.}

\def\cvprPaperID{5896} 
\def\confName{CVPR}
\def\confYear{2023}

\begin{document}

\title{DOLCE: A Model-Based Probabilistic Diffusion Framework \\ for Limited-Angle CT Reconstruction}

\author{Jiaming Liu$^{\footnotesize 1}$\space\space\space\space\space\space\space\space\space\space  
Rushil Anirudh$^{\footnotesize 2}$\space\space\space\space\space\space\space\space\space\space  
Jayaraman J. Thiagarajan$^{\footnotesize2}$\space\space\space\space\space\space\space\space\space\space  
Stewart He$^{\footnotesize 2}$\\ 
K. Aditya Mohan$^{\footnotesize 2}$\space\space\space\space\space\space\space\space\space\space 
Ulugbek S. Kamilov$^{\footnotesize 1}\thanks{Corresponding authors.}$\space\space\space\space\space\space\space\space\space
Hyojin Kim$^{\footnotesize 2}$$^\ast$\\
$^{\footnotesize 1}$Washington University in St.~Louis\space\space\space
$^{\footnotesize 2}$Lawrence Livermore National Laboratory\\
{\small \tt \{jiaming.liu, kamilov\}@wustl.edu,} {\small \tt \{anirudh1, jayaramanthi1, he6, mohan3, hkim\}@llnl.gov} 
}

\maketitle

\begin{abstract}
Limited-Angle Computed Tomography (LACT) is a non-destructive evaluation technique used in a variety of applications ranging from security to medicine. The limited angle coverage in LACT is often a dominant source of severe artifacts in the reconstructed images, making it a challenging inverse problem. We present DOLCE, a new deep model-based framework for LACT that uses a conditional diffusion model as an image prior. Diffusion models are a recent class of deep generative models that are relatively easy to train due to their implementation as image denoisers. DOLCE can form high-quality images from severely under-sampled data by integrating data-consistency updates with the sampling updates of a diffusion model, which is conditioned on the transformed limited-angle data. We show through extensive experimentation on several challenging real LACT datasets that, the same pre-trained DOLCE model achieves the SOTA performance on drastically different types of images. Additionally, we show that, unlike standard LACT reconstruction methods, DOLCE naturally enables the quantification of the reconstruction uncertainty by generating multiple samples consistent with the measured data. 
\end{abstract}

\section{Introduction}
\label{sec:intro}

\begin{figure}[t!]
	\centering
	\includegraphics[width=\linewidth]{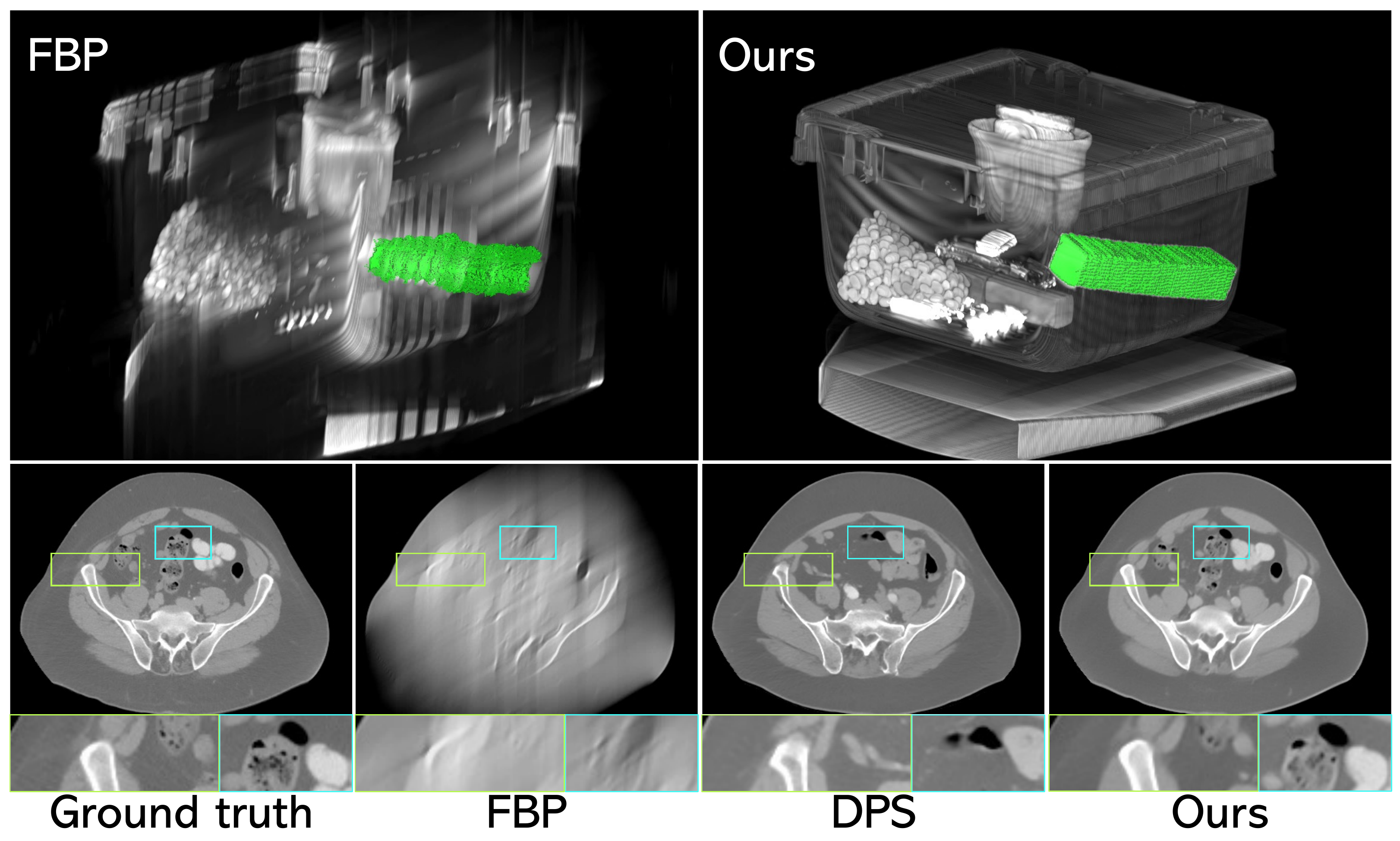}
	\caption{We show that the same pre-trained DOLCE model can reconstruct distinct CT images such as checked-in luggage~\cite{TO4.etal2014} and human body~\cite{mccollough2016tu}. \emph{Top:} 3D rendering of a luggage from its 2D slices reconstructed using DOLCE on the limited-angle data containing just one-third of the views (0-60$^{\circ}$). Note how our method preserves the 3D edges, enabling a successful recovery of the object geometries.~\emph{Bottom:} Comparison of DOLCE on a medical dataset with DPS, which is a SOTA method for solving imaging inverse problems using unconditional diffusion models~\cite{Chung.etal2022a}. See Section~\ref{Sec:Experiments} for the complete set of experimental results.
    }
    \label{Fig:fig1}
    \vspace{-0.5em}
\end{figure}

Computed Tomography (CT) is one of the most widely-used imaging modalities with applications in medical diagnosis, industrial non-destructive testing, and security~\cite{Wang.etal2018, De.etal2014, Wang.etal2008, Wells.etal2012}. In a typical parallel-beam CT imaging system, the x-ray measurements obtained from all viewing angles are combined to reconstruct a cross-sectional image of a 3D object~\cite{Kak.etal2001}. Conventional reconstruction methods such as Filtered Back Projection (FBP) can produce high-quality CT images given a complete set of projection data, but completely fail under more ill-posed scenarios such as  \emph{Limited-Angle CT (LACT)}, where projections from only a limited range of angles can be acquired (\emph{i.e.}, $0 \leq \theta \leq \theta_{\text{max}}$ with $\theta_{\text{max}}<\pi$)~\cite{Bachar.etal2007, Quinto.etal2008, Hyv.etal2010, Cho.etal2013, Mohan.etal2015}. A typical solution to this inverse problem is model-based optimization that integrates a forward model characterizing the imaging system and a regularizer imposing priors on the unknown image. While there has been significant progress in algorithms that leverage sophisticated image priors (\emph{e.g.}, transform-domain sparsity, self-similarity, and learned dictionaries)~\cite{Figueiredo.Nowak2001, Elad.Aharon2006, Kudo.etal2013image, Danielyan.etal2012}, the focus in the area has recently shifted to deep learning (DL). 

\medskip\noindent
\textbf{Deep Learning for CT:} A traditional DL reconstruction involves training a convolutional neural network (CNN) architecture, such as U-Net~\cite{Ronneberger.etal2015}, to directly perform a regularized inversion of the forward model by exploiting redundancies in the training data~\cite{DJin.etal2017, Huang.etal2017b, Gupta.etal2018, han.etal2018, Ye.etal2018, Anirudh.etal2018, Zhang.etal2021d,Zhang.etal2020b}. Model-based DL (MBDL) is another popular reconstruction strategy that seeks to explicitly use the knowledge of the forward model by integrating a CNN into a model-based algorithm. Popular MBDL frameworks include Plug-and-Play Priors (PnP)~\cite{Venkatakrishnan.etal2013, Romano.etal2017}, which use pre-trained deep denoisers as image priors~\cite{Sreehari.etal2016, Metzler.etal2018, Zhang.etal2021b}, and Deep Unfolding~\cite{Gilton.etal2020, Adler.etal2018, Liu.etal2021, Gao.etal2022, Zhou.etal2021, Hu.etal2022}, which interpret the iterations of a model-based algorithm as layers of a CNN to perform end-to-end supervised training. Other DL strategies used in CT reconstruction include dual-domain learning~\cite{Usman.etal2019, Zhou.etal2022}, deep internal learning~\cite{Zang.etal2021, Ruckert.etal2022, Vasconcelos.etal2022}, and measurement synthesis learning~\cite{lee.etal2017, Lee.etal2019, Sun.etal2021b}. Despite the rich literature on tomographic imaging, the reconstruction of high-quality images with sharp edges remains a well-known challenge, particularly when the acquired data is missing a large-range of angles (\emph{i.e.}, $\theta_{\text{max}} \leq 90^{\circ}$). Furthermore, most prior work in the area has focused on methods that can only produce point estimates without any quantification of the reconstruction uncertainty, which can be essential in critical applications such as healthcare or security.

\begin{figure*}[t!]
	\centering
	\includegraphics[width=0.99\textwidth]{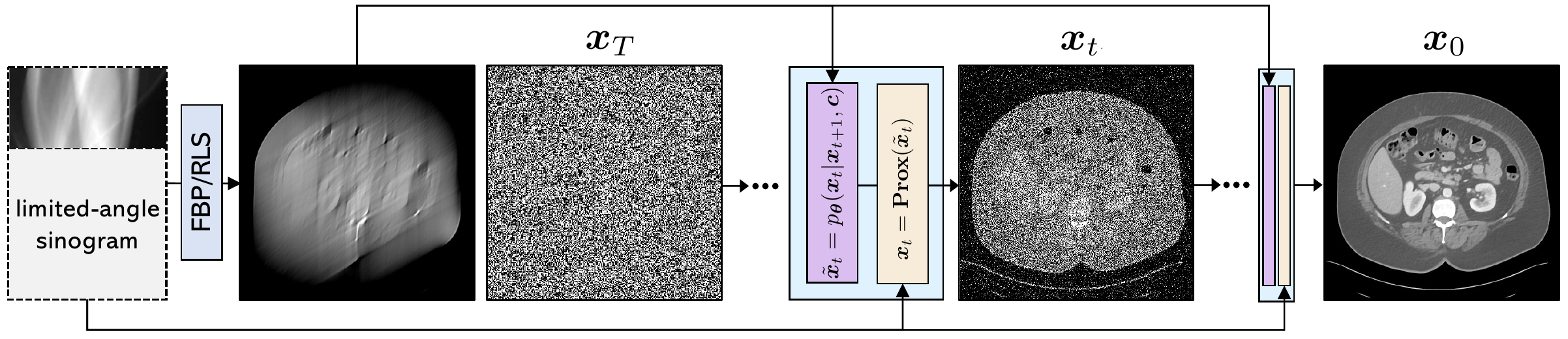}
	\caption{
    Overview of the proposed approach. Starting from the Gaussian noise $\xbm_{T}$, we sample an image $\xbm_{0}$ from the posterior by solving the reverse process of conditional denoising diffusion model, alternating between the denoising-update step and the data-consistency step.}
	\label{Fig:fig2}
\end{figure*}

\medskip\noindent
\textbf{Proposed Work:} We present \emph{\textbf{D}iffusion Pr\textbf{o}babilistic \textbf{L}imited-Angle \textbf{C}T R\textbf{e}construction (DOLCE)}, a conditional generative model for LACT, which can generate multiple diverse, yet high-quality, reconstructions from a given limited-angle data. Inspired by the recent successes of Denoising Diffusion Probabilistic Models (DDPM)~\cite{Saharia.etal2022a, Dhariwal.etal2021} and denoising score matching~\cite{Song.etal2019, Song.etal2020a}, we design DOLCE as a ``repeated-refinement'' conditional diffusion model. Specifically, DOLCE trains a stochastic sampler conditioned on noisy seed reconstructions obtained using transformed limited-angle sinograms. %
To boost the imaging quality further, DOLCE imposes an additional data-consistency step at every iteration after the sampling-update step. DOLCE can thus be viewed as a method for transforming a standard normal distribution into an empirical data distribution through a sequence of refinement steps, while integrating physical forward models and learned stochastic samplers (see Fig.~\ref{Fig:fig2}). 

We demonstrate several unique features of DOLCE compared to the prior work through extensive experimentation on two real-world LACT datasets. We first show that, on both applications, DOLCE achieves the \emph{state-of-the-art (SOTA)} performance by directly producing high-resolution $512 \times 512$ images across a range of limited-angle scenarios ($\theta_{\text{max}}\in\{60^{\circ}, 90^{\circ}, 120^{\circ}\}$).  Next, we make an interesting finding that the same pre-trained DOLCE model can be effective on LACT from significantly different data distributions, such as images of human body and of checked-in luggage, enabling highly generalizable CT reconstruction networks for the first time.  Finally, we show how the diverse realizations produced by DOLCE (from a given sinogram) can enable meaningful uncertainty quantification~\cite{Kendall.etal2017}. Notably, we find the variances estimated by DOLCE to be well calibrated, \emph{i.e.}, consistent with the true reconstruction errors. In short, DOLCE is the first model-based probabilistic diffusion framework for LACT that achieves SOTA performance and enables systematic uncertainty characterization.

\medskip
\noindent Our main contributions can be summarized as follows:
\begin{enumerate}
  \item We propose DOLCE as the first conditional diffusion model for the recovery of high-quality CT images from limited-angle sinograms. 
  
  \item We show that DOLCE is effective across two real-world datasets: checked-in luggage and medical-image datasets. DOLCE achieves a PSNR improvement of at least 3 dB over ILVR~\cite{Choi.etal2021} and DPS~\cite{Chung.etal2022a}, two SOTA diffusion models for inverse problems.
  
  \item We use DOLCE to provide uncertainty maps for the reconstructed LACT images. The uncertainty estimates are reflective of the true reconstruction errors.
  
  \item Using a 3D segmentation experiment, we show the effectiveness of DOLCE in recovering the geometric structure and sharp edges in high-resolution images, even in severely ill-posed settings.

\end{enumerate}

\section{Related Work}
\label{sec:related}
\noindent
\textbf{Tomographic Image Reconstruction.} Traditional analytic algorithms such as FBP are commonly used for CT reconstruction. However, FBP produces inaccurate reconstructions with noise and artifacts when the imaging conditions are highly ill-posed such as in limited angle or sparse-views scenarios. Iterative methods are a popular alternative for tomographic reconstruction. Earlier works such as Algebraic Reconstruction Techniques (ART)~\cite{Gordon.etal1970}, solve a discrete formulation of the reconstruction problem.  These approaches can be combined with regularizers in an optimization framework, resulting in model-based iterative reconstruction (MBIR) algorithms~\cite{Mohan.etal2015, Huang.etal2018, Xu.etal2020b, Venkatakrishnan.etal2021, Bouman.etal2022, HAADF_Venkat2013}. MBIR optimizes the reconstruction solution such that it best fits to the forward model, which captures the measurement physics and noise statistics, and a prior model for the object. 

Recent DL-based methods adopt an end-to-end approach where a deep network architecture is trained in a supervised fashion to directly produce a point estimate~\cite{DJin.etal2017, Huang.etal2017b, Anirudh.etal2018, fu.etal2019, Zhang.etal2020b, hu.etal2021, Bubba.etal2021, Zhang.etal2021d, Gao.etal2022}. For example, ~\cite{Gao.etal2022, Cheng.etal2020, Zhou.etal2021, Liu.etal2021, Hu.etal2022} propose to unfold an iterative algorithm and train it end-to-end as a deep neural network. This enables integration of the physical information into the architecture in the form of data-consistency blocks that are combined with trainable CNN regularizers. Deep internal learning methods are alternates for tomographic reconstruction that explore the internal information of the test signal for learning a neural network prior without using any external data~\cite{Gadelha.etal2019, Zang.etal2021, Ruckert.etal2022, Baguer.etal2020, Wu.etal2022}. A related family of denoising-driven approaches known as PnP algorithms represents alternative to traditional DL methods by combining iterative model-based algorithms with deep denoisers as priors and have been shown to be effective in various forms of tomographic imaging~\cite{Majee.etal2021, Wei.etal2022, Sun2019b, Liu.etal2022}.

\noindent
\textbf{Diffusion Models in Imaging.} Denoising diffusion models~\cite{Hao.etal2020, Dhariwal.etal2021, Kingma.etal2021} and score-based models~\cite{Song.etal2019, Song.etal2020a, Song.etal2020b} are two related classes of generative models that were shown to achieve the SOTA performance in unconditional image generation. Despite being discovered independently, both classes are often referred to as~\emph{diffusion models} due to their similarity~\cite{Huang.etal2021, Kingma.etal2021}. Diffusion models are trained to model the Markov transition from a simple distribution to the data distribution, enabling the generation of samples through sequential stochastic transitions. Apart from unconditional image generation, diffusion models have recently been applied to conditional image generation, including super-resolution~\cite{Choi.etal2021, Saharia.etal2022a, Chung.etal2022b}, sparse-CT reconstruction~\cite{Song.etal2022, Chung.etal2022b}, MRI reconstruction~\cite{Chung.etal2022b, Chung.etal2022c, Xie.etal2022}, and phase-retrieval~\cite{Chung.etal2022a}. For example, one line of work has focused on designing diffusion models suitable for specific image reconstruction tasks~\cite{Saharia.etal2022a, Xia.etal2022, Chung.etal2022c}. Another line of work has focused on keeping the training of a diffusion model intact, and only modify the inference procedure to enable sampling from a conditional distribution~\cite{Choi.etal2021, Kawar.etal2022, Chung.etal2022a, Song.etal2022, Chung.etal2022b, Chung.etal2022, Lugmayr.etal2022}. These methods can be thought of as solving different image reconstruction tasks by leveraging the learned score function as a generative prior of the data distribution. However, for the severely ill-posed LACT reconstruction, the current SOTA diffusion models often fail to generate images with desired semantics and accurate details (see Section~\ref{Sec:Experiments}). The proposed DOLCE method addresses this issue by integrating conditional learning and model-based inference for SOTA reconstruction in LACT.





\section{Preliminaries}

\noindent
\textbf{Inverse Problems.} The problem of LACT reconstruction can be formulated as a linear inverse problem involving the recovery of an image $\xbm \in \R^n$ from incomplete measurements $\ybm = \Abm\xbm$, where $\Abm \in \R^{m \times n}$ is the measurement operator modeling the observation process. Recovering $\xbm$ from $\ybm$ in LACT is highly ill-posed, often requiring addition assumptions on the unknown $\xbm$. From the Bayesian statistical perspective, the estimation can be viewed as sampling from the posterior distribution $p(\xbm|\ybm)$. One can also compute point estimates of $\xbm$ using the maximum-a-posteriori probability (MAP) $\argmax{p(\xbm|\ybm)}$ or minimum mean square error (MMSE)~$\E[\xbm|\ybm]$ estimators.

\medskip\noindent
\textbf{Denoising Diffusion Probabilistic Models.} DDPM refers to generative models that learn a target data distribution from samples~\cite{Hao.etal2020, Sohl.etal2015}. DDPM consists of two Markov processes: the fixed forward process and the learning-based reverse process. The forward process starts from a sample of a clean image $\xbm_0\sim q(\xbm_0)$ and gradually adds Gaussian noise according to the following transition probability:
\begin{equation}
    \label{eq:ForwardDiffusion}
    q(\xbm_t|\xbm_{t-1}) := \Ncal(\xbm_t; \sqrt{1- \beta_{t}} \xbm_{t-1}, \beta_t \Ibf),
\end{equation}
where $\Ncal(\cdot)$ denotes the Gaussian pdf, $\beta_{1:T}$ refers to a variance schedule subject to $\beta_t \in (0,1)$  for all $t = 1,\cdots,T$. The latent variables $\xbm_{1:T}$ have the same dimensionality as the original image
sample $\xbm_0\in\R^n$, and latent $\xbm_T$ is nearly an isotropic Gaussian distribution for large enough $T$ and a properly selected $\beta_t$ schedule. By parameter change of $\alpha_t := 1- \beta_t$ and $\Bar{\alpha}_t = \Pi_{s=1}^t \alpha_s$, we can write $\xbm_t$ as a linear combination of noise $\epsilonbm$ and $\xbm_0$ 
\begin{equation}
\label{eq:Diff2}  
\xbm_t = \sqrt{\bar{\alpha}_t}\xbm_0 + \sqrt{1-\bar{\alpha}_t} \epsilonbm,
\end{equation}
where $\epsilonbm \sim \Ncal (0, \Ibf)$. This allows a closed-form expression for the marginal distribution for sampling $\xbm_t$ given $\xbm_0$
\begin{equation}
    \label{eq:sampleNois}
    q(\xbm_t|\xbm_{0}) := \Ncal(\xbm_t; \sqrt{\Bar{\alpha}_t} \xbm_{0}, ( 1- \Bar{\alpha}_t) \Ibf).
\end{equation}

\noindent
\textbf{Improved Reverse Process.} Since the reverse process $q(\xbm_{t-1}|\xbm_t)$ depends on the entire data distribution and is not tractable, we can learn the parameterized Gaussian transitions $p_{\thetabm}(\xbm_{t-1}|\xbm_{t})$ using a neural network as follows:  
\begin{equation}
    \label{eq:reverse1}
    p_{\thetabm}(\xbm_{t-1}|\xbm_{t}) = \Ncal(\xbm_{t-1}; \mu_{\thetabm}(\xbm_t, t), \sigma_{t}^2\Ibf),
\end{equation}
where $\mu_{\thetabm}(\xbm_t, t)$ refers to the learned mean. It is worth noting that originally Ho~\emph{et al.}~\cite{Hao.etal2020} set the variance $\sigma_{t}$ to a fixed constant value. However, subsequent works~\cite{Nichol.etal2021, Dhariwal.etal2021} proved the improved generation efficiency by using learned variance $\sigma_t^2\defn\sigma^2_{\thetabm}(\xbm_t, t)$, which we also adopt. In particular, the variances $\sigma_{\thetabm}(\xbm_t,t)\defn \exp(v\log\beta_t + (1-v)\log\tilde{\beta}_t)$, correspond to the output of the neural network and $\tilde{\beta}_t$ refers to the lower bounds for the reverse process variances~\cite{Hao.etal2020}. We use a single neural network with two separate output heads to estimate the mean and variance of this Gaussian distribution jointly. Practically, one can relate $\xbm_t$ and $\xbm_0$ via Equation~\eqref{eq:Diff2} and~\eqref{eq:sampleNois} by decomposing $\mu_{\thetabm}$ into a linear combination of $\xbm_{t}$ and the noise approximation $\epsilon_{\thetabm}$. More specifically, we have $\xbm_t=\sqrt{\bar{\alpha}}\xbm_0 + \sqrt{(1-\bar{\alpha})}\epsilonbm$ for $\epsilonbm\sim\Ncal(0,\Ibf)$ and can train the network $\epsilon_{\thetabm}$ as a denoiser to predict $\epsilonbm$.
During sampling, we can use simple substitution to derive $\mu_{\thetabm}(\xbm_t,t)$ from network prediction $\epsilon_{\thetabm}(\xbm_t,t)$
\begin{equation}
    \label{eq:generalsamplingrule2}
    \xbm_{t-1} = \frac{1}{\sqrt{\alpha_t}}\bigg(\xbm_t - \frac{1 - \alpha_t}{\sqrt{1- \Bar{\alpha}_t}} \epsilon_{\thetabm}(\xbm_t, t)\bigg) + \sigma_t \zbm, 
\end{equation}
where $\zbm \sim \Ncal(0,\Ibf)$. Since the model learns the reverse Markov Chain running backward in time from $\xbm_T$ to $\xbm_0$, estimating clean image $\xbm_0$ from partially noisy image $\xbm_t$, we refer to this as the ~\emph{reverse process}.
\label{sec:Preliminaries}
\section{Proposed Approach: DOLCE}
In this section, we present our proposed approach for LACT, and describe the training and testing strategies. An overview of DOLCE is provided in
Fig.~\ref{Fig:fig2}. Our goal here is to reconstruct full-view images sampled from the conditional distribution $p(\xbm_0|\cbm)$, where the condition $\cbm$ is obtained from a limited angle sinogram. Specifically, we make the neural network accept $\cbm$ as the conditioning input. Note that while related ideas have been considered in other applications, such as  image blurring~\cite{Whang.etal2022} and super-resolution~\cite{Saharia.etal2022a}, our work is the first to adopt conditional sampling for CT reconstruction. This way, the iterative denoising procedure becomes dependent on $\cbm$ and the conditional diffusion model can generate a target image $\xbm_0$ in $T$ refinement steps. Starting from step $T$, each Markov transition under the condition $\cbm$ is approximated as follows:
\begin{equation}
\begin{aligned}
    \label{eq:cond-reverse1}
    &\quad\quad p_{\thetabm}(\xbm_{0:T}|\cbm) = p(\xbm_T)\prod_{t=1}^{T}p_{\thetabm}(\xbm_{t-1}|\xbm_t,\cbm),\\
    &p_{\thetabm}(\xbm_{t-1}|\xbm_t, \cbm) = \Ncal(\xbm_{t-1};\mu_{\thetabm}(\xbm_t,\cbm,t),\, \textbf{diag}(\bm{\sigma}_t^2)) ,   
\end{aligned}    
\end{equation}
where $\xbm_T$ is sampled from the normal distribution $p(\xbm_T)\sim\Ncal(0,\Ibf)$, and we use $\bm{\sigma}_t^2 \defn \sigma^2_{\thetabm}(\xbm_t, \cbm,t)$ to denote the learned variances. Similar to the reverse process of unconditional model, the inference process $p_{\thetabm}(\xbm_{t-1}|\xbm_{t},\cbm)$ is learned using a neural network that takes the conditional data $\cbm$ as an additional input. 
\begin{algorithm}[t]
\caption{DOLCE Iterative Refinement}\label{alg:redbls}
\begin{algorithmic}[1]
\State \textbf{Input:} $\tilde\epsilon_{\thetabm}$: Adjusted denoiser network, $\cbm$: Conditional inputs image (FBP or RLS), $g$: Data-fidelity;
$\gamma_t> 0$;
\State \textbf{Output:} Restored image $\xbm_0$
\State \text{Sample} $\xbm_{T} \sim \Ncal({\zerobm, \Ibf})$   \Comment{Run diffusion sampling}
\For{$t = T,\dots,1$}
\State $\zbm\sim\Ncal(0,\Ibf)$
\State $\xbmtilde_{t-1} = \frac{1}{\sqrt{\alpha_t}}(\xbm_t - \frac{1 - \alpha_t}{\sqrt{1- \Bar{\alpha}_t}} \tilde\epsilon_{\thetabm}(\xbm_t, \cbm, t)) + \bm{\sigma}_t \cdot \zbm,$
\State $\xbm_{t-1}=\prox_{\gamma_t g}{(\xbmtilde_{t-1})}$ \Comment{Proximal operator}
\EndFor
\State \textbf{return:} $\xbm_0$
\end{algorithmic}
\end{algorithm}
\subsection{Optimizing the Conditional Denoising Network}
While it would be possible to impose the condition $\cbm$ directly from the measurement domain, we find that using a low-fidelity reconstruction, from any standard inversion technique, to define $\cbm$ greatly simplifies the learning. Similar approaches are routinely used in traditional full-view CT reconstruction~\cite{DJin.etal2017, han.etal2018, Liu.etal2021}.
Popular choices for standard inversion include FBP and the regularized least squares (RLS).  Note that our approach is generic enough to support the use of other condition specifications as well. In practice, the choice is made based on both the inversion quality and computational efficiency. For example, RLS inversion is known to be time-efficient, due to efficient GPU implementations, and can produce better quality reconstructions. Hence, we concatenate $\xbm_t$ with reconstruction from RLS along the channel dimension to condition the model, leading to the training objective:
\begin{equation}
    \label{eq:condloss}
    L_{\text{base}} = \E_{\xbm_0,\cbm,\epsilonbm, t\sim[1,T]}\left[\|\epsilonbm - \epsilon_{\thetabm}(\xbm_t, \cbm, t)\|^2\right],
\end{equation}where $\cbm\in\R^n$ has the same dimension as latent variables $\xbm_{1:T}$. Similar to~\cite{Nichol.etal2021}, we did not apply any training constraints on $\sigma_{\thetabm}(\xbm_t,\cbm,t)$, and we did not observe any noticeable performance drop, suggesting that the bounds for $\sigma_{\thetabm}(\xbm_t,\cbm,t)$ are expressive enough. 

In order to improve the generation flexibility, we jointly train a single diffusion model on conditional and unconditional objectives by randomly dropping $\cbm$ during training (\eg, $p_{\text{uncond}}=0.2$), similar to the \emph{classifier free guidance}~\cite{Ho.etal2021, Saharia.etal2022b}. Hence, the sampling is performed using the adjusted noise prediction:
\begin{equation}
    \label{eq:reverse4}
    \tilde\epsilon_{\thetabm}(\xbm_t,\cbm,t)=\lambda\epsilon_{\thetabm}(\xbm_t,\cbm,t) + (1-\lambda)\epsilon_{\thetabm}(\xbm_t,t),
\end{equation}where $\lambda>0$ is the trade-off parameter, and $\epsilon_{\thetabm}(\xbm_t,t)$ is the unconditional $\epsilonbm$-prediction. For example, setting $\lambda=1$ disables the unconditional guidance, while
increasing $\lambda > 1$ strengthens the effect of conditional $\epsilonbm$-prediction. 

\subsection{Model Based Iterative Refinement}
It is well known that sinograms have certain consistency conditions that are hard to enforce entirely within the neural network. As such, given the trained conditional diffusion model, we propose to directly enforce consistency with the limited-angle sinogram $\ybm$. This is done during inference by including an additional step to the denoising iteration update conditioned on the FBP or RLS. Similar to the reverse process~\eqref{eq:generalsamplingrule2} of the unconditional diffusion model, each iteration of iterative refinement under our adjusted denoising model takes the form:
\begin{equation}
    \label{eq:generalsamplingrule1}
    \xbmtilde_{t-1} = \frac{1}{\sqrt{\alpha_t}}\bigg(\xbm_t - \frac{1 - \alpha_t}{\sqrt{1- \Bar{\alpha}_t}} \tilde\epsilon_{\thetabm}(\xbm_t, \cbm, t)\bigg) + \bm{\sigma}_t \cdot \zbm, 
\end{equation}
where $\zbm \sim \Ncal(0,\Ibf)$. This resembles one step of Langevin
dynamics with $\tilde\epsilon_{\thetabm}$ providing an estimate of the gradient of the data log-density. Then, the data consistency mapping under $\ell_2$-norm loss is promoted by solving a proximal optimization~\cite{Moreau1965} step:
\begin{equation}
    \label{eq:proxrule}
    \xbm_{t-1} = \argmin_{\zbm\in\R^n}\left\{\|\zbm - \xbmtilde_{t-1}\|_2^2 + \gamma_t\|\Abm\zbm - \ybm\|_2^2\right\},
\end{equation}
where the parameter $\gamma_{t} > 0$ at each step balances the importance of the data consistency $\|\Abm\zbm-\ybm\|_2^2$. Since our implementation of the forward and backward projection uses GPU accelerated backend~\footnote{Implementation using the Pytorch’s Custom C++ and CUDA extensions}, the sub-problem~\eqref{eq:proxrule} can be efficiently solved with any gradient-based method, \emph{e.g.}, conjugate-gradient~\cite{Hestenes.etal1952} or accelerated-gradient methods~\cite{Beck.Teboulle2009}.

\noindent
\textbf{Sample Average.} Similar to~\cite{Whang.etal2022}, since our model is designed to sample from the target posterior $p(\xbm|\ybm)$, we can average multiple samples from our method to approximate the conditional mean $\E[\xbm|\ybm]$. Hence, we also report results averaged over multiple samples, denoted as ``DOLCE-SA''. 

\subsection{Model Architecture and Sampling Schedules}
The network architecture within DOLCE is similar to the U-Net in~\emph{guided diffusion}~\cite{Dhariwal.etal2021}, with self-attention and modifications adapted from~\cite{Song.etal2020b}, where the original DDPM residual blocks are replaced with residual blocks from BigGAN~\cite{Brock.etal2018}, and the skip connections are re-scaled with $1/\sqrt{2}$ for faster training convergence. In addition, we add time-embedding into the attention bottle block, and we increase the number of residual blocks at lower-resolution in order to increase the model capacity through more model parameters. 

For our training noise schedule we set $T=2000$, and the variance $\beta_t$'s are uniformly spaced. We also experimented with a~\emph{cosine} noise schedule proposed in IDDPM~\cite{Dhariwal.etal2021} during training, but observed similar image reconstruction quality. At inference time, early diffusion models~\cite{Hao.etal2020, Song.etal2020b} require the same number of diffusion steps ($T$) as training, making generation slow, especially for high-resolution images. For a more efficient generation (inference), we use $K\in[1,T)$ evenly spaced real numbers, and then round each resulting number to the nearest integer following~\cite{Dhariwal.etal2021}. In addition, we run a grid search over the hyperparameters of the proximal step and the rescheduling time step $K$ for the best peak signal-to-noise-ratio score (PSNR). This inference-time hyperparameter tuning is cheap as it does not involve retraining or fine-tuning the model itself.

\begin{figure*}[t!]
	\centering
	\includegraphics[width=0.92\textwidth]{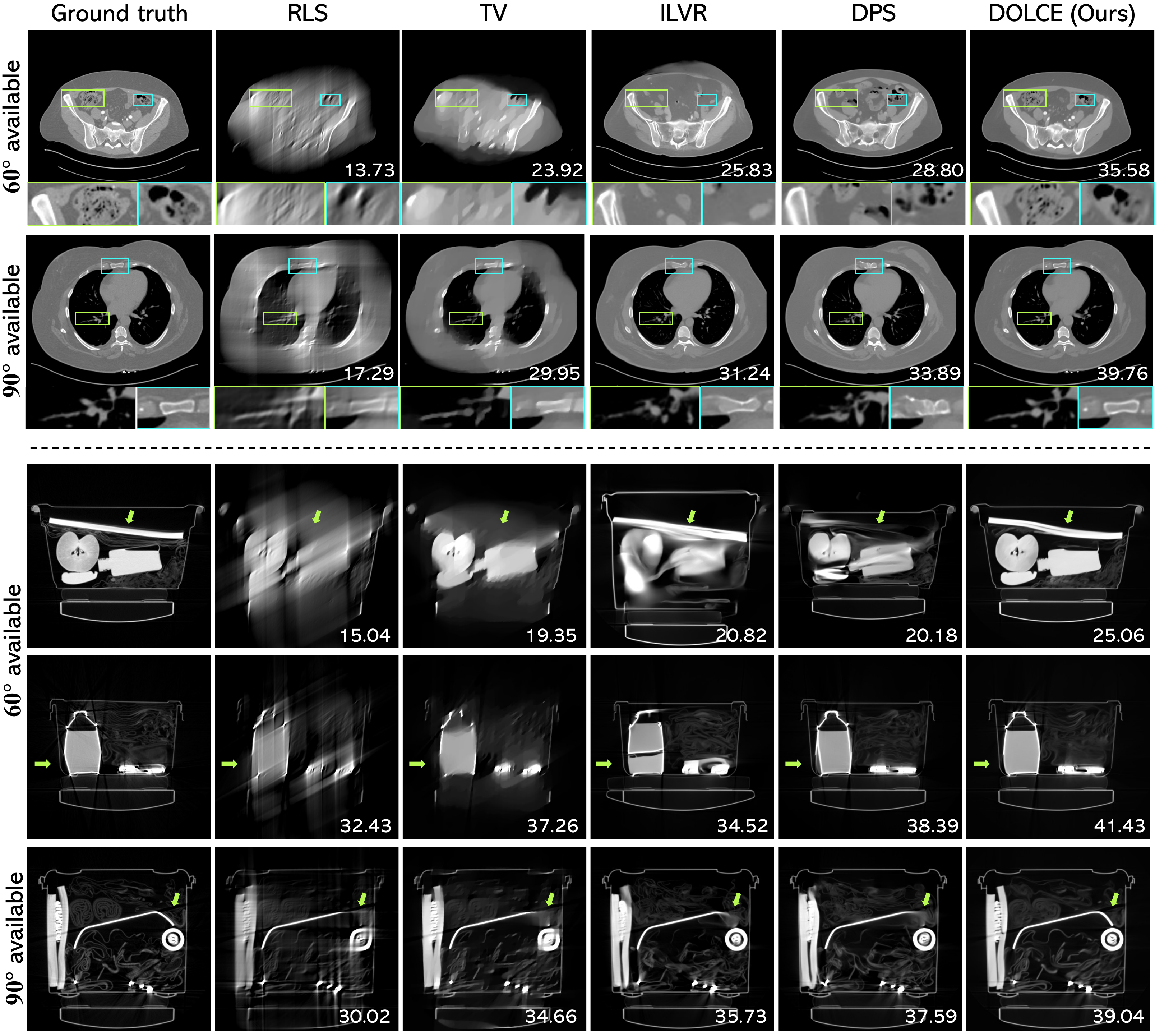}
	\caption{Visual evaluation of limited angle tomographic reconstruction in body CT scan (\textbf{top}) and checked-in luggage (\textbf{bottom}), where the input measurements are captured respectively from an angular coverage of 60$^{\circ}$ and 90$^{\circ}$, respectively. PSNR (dB) is indicated at bottom for each reconstruction, measured against the ground truth. Note the remarkable accuracy of DOLCE reconstructions that preserve fine image details. See Table~\ref{table:results1} and Table~\ref{table:results2} for quantitative comparisons with additional baselines. Images are normalized for better visualization.
	}
	\label{Fig:fig3}
\end{figure*}
\section{Experiments}
\label{Sec:Experiments}

\subsection{Datasets}
\noindent
\textbf{Checked-in Luggage Dataset.}
The luggage dataset is collected using an Imatron electron-beam medical scanner – a device similar to those found in transportation security systems, provided by the DHS ALERT Center of Excellence at Northeastern University~\cite{TO4.etal2014} for the development and testing of Automatic Threat Recognition (ATR) systems. The dataset is comprised of $190$ bags, with roughly $300$ slices per bag on an average. In total, the dataset consists of $50$K full view sinograms along with their corresponding FBP reconstructions. The image matrix is resampled to be $512\times512$, and correspondingly the sinograms are subsampled to be of size 720×512. This corresponds to views obtained at every $0.25^\circ$ uniformly sampled from $180^\circ$. We repurpose this dataset for generating CT reconstructions from sinograms. We split the bags into a training set of $165$ bags and a test set with the rest, corresponding to about $40$K for training and $10$K for testing. The bags contain a variety of everyday objects, including clothes, food, electronics etc., that are  arranged in random configurations.

\begin{table}[h!]
  	\centering
  	\resizebox{\columnwidth}{!}
  	{\begin{tabular}{lcccccc}
	\toprule
 	\multicolumn{1}{l}{\textbf{Metric}} & \multicolumn{3}{c}{\bf PSNR $\,\uparrow$} & \multicolumn{3}{c}{\bf SSIM $\,\uparrow$}\\
  	\cmidrule{2-4} \cmidrule{5-7} 
\textbf{Angle}           & \bf 60$^\circ$ & \bf 90$^\circ$ & \bf 120$^\circ$  & \bf 60$^\circ$ & \bf 90$^\circ$ & \bf 120$^\circ$\\ 	
 	\hline\hline
FBP                                      & 15.17 & 17.51 & 21.20  & 0.464 & 0.540 & 0.601\\
RLS                                      & 22.75 & 26.26 & 30.47  & 0.698 & 0.832 & 0.887\\
TV~\cite{Beck.Teboulle2009}                & 25.60 & 30.27 & 36.33  & 0.791 & 0.907 & 0.956\\
U-Net~\cite{DJin.etal2017}             & 26.86 & 31.31 & 38.61  & 0.852 & 0.932 & 0.966\\
DPIR~\cite{Zhang.etal2021b}                & 26.22 & 31.25 & 37.60  & 0.849 & 0.930 & 0.951\\
ILVR~\cite{Choi.etal2021}                 & 28.63 & 33.34 & 37.68  & 0.861 & 0.931 & 0.955\\
DPS~\cite{Chung.etal2022a}               & 28.97 & 33.45 & 37.92  & 0.897 & 0.937 & 0.959\\
\hline
DOLCE                                    & \colorbox{lightblue}{\makebox(24,4){35.11}} & \colorbox{lightblue}{\makebox(24,4){39.04}} & \colorbox{lightblue}{\makebox(24,4){42.16}}  & \colorbox{lightblue}{\makebox(24,4){0.941}} & \colorbox{lightblue}{\makebox(24,4){0.959}} & \colorbox{lightblue}{\makebox(24,4){0.971}}\\
DOLCE-SA                                 & \colorbox{lightgreen}{\makebox(24,4){\bf35.58}} & \colorbox{lightgreen}{\makebox(24,4){\bf39.61}} & \colorbox{lightgreen}{\makebox(24,4){\bf42.84}}  & \colorbox{lightgreen}{\makebox(24,4){\bf0.946}} & \colorbox{lightgreen}{\makebox(24,4){\bf0.963}} & \colorbox{lightgreen}{\makebox(24,4){\bf0.975}}\\
    \hline
	\end{tabular}}
	\caption{
	Average PSNR and SSIM results for several methods on human body CT dataset. \colorbox{lightgreen}{\makebox(45,6){\textbf{Best values}}}~and~\colorbox{lightblue}{\makebox(70,6){second-best values}} for each metric are color-coded.}
	\label{table:results1}
	\vspace{-0.5em}
\end{table}
\vspace{-0.0em}
\noindent
\textbf{Body CT Scan Datasets.} We additionally use Kidney CT scans of 210 patients from the publicly available dataset~\emph{2019 Kidney and Kidney Tumor Segmentation Challenge (C4KC-KiTS)}~\cite{C4KC_Heller2019}. The collection contains 406 scans, where each patient has 1-3 scans. Each 3D scan consists of about $92 \sim 812$ 2D slices covering a range of anatomical regions from chest to pelvis, resulted in about $70$K slices in total. We choose $60$K 2D slices of size $512 \times 512$ corresponding to 190 patients to train the models. The test images correspond to $10$K slices randomly selected from the remaining patients.

\subsection{Training details and parameters}
We train and evaluate the models with Pytorch using Tesla V100 GPUs with 16GB memory. To show the effectiveness of our conditional diffusion model, we train a single DOLCE model on the luggage and body CT dataset jointly, by minimizing the loss in Eq.~\eqref{eq:condloss}. We rescale each dataset globally to make them have the same intensity range, but we do not perform any normalization on those images. As baselines for comparison, we also train individual models on luggage and body CT dataset. For both two datasets, FBP and RLS reconstructions are obtained using publicly available CT reconstruction tools such as LTT~\cite{LTT} and TomoPy \cite{gursoy2014tomopy}. During training, we randomly select FBP or RLS reconstructed using $\theta_{\text{max}}\in\{60^{\circ}, 90^{\circ}, 120^{\circ}\}$ as the conditional input, so that the models can handle multiple scenarios. The FBP or RLS is normalized to intensity range of $[0,1]$ for better performance and stable training. We also train two unconditional diffusion models on each dataset and one on the joint dataset as additional baselines. Due to GPU memory constraints, we train all diffusion models in half precision (\texttt{float16}) with a batch-size of $256$. We use the Adam optimizer with a fixed learning rate of $1.5\times 10^{-4}$ and a dropout rate of $0.2$ for each model. We do not perform any checkpoint selection on our models and simply select the latest checkpoint. It takes about two days to obtain a DOLCE model.

\subsection{Quantitative and Qualitative Results}
Table~\ref{table:results1} and Table~\ref{table:results2} show average PSNR and SSIM~\cite{Wang.etal2004} results of several methods for 150 randomly chosen slices from each test set, respectively. The compared methods include FBP, RLS, TV~\cite{Beck.Teboulle2009}, U-Net~\cite{DJin.etal2017}, CTNet~\cite{Anirudh.etal2018}, DPIR~\cite{Zhang.etal2021b}, ILVR~\cite{Choi.etal2021}, and DPS~\cite{Chung.etal2022a}. Note that CTNet is a method specifically designed for luggage dataset to reconstruct directly from sinograms. We observed that making CTNet perform well on other datasets requires dedicated fine-tuning so we omit its results on medical dataset for fair comparison. U-Net corresponds to our own implementation of the architecture used in the FBPConvNet~\cite{DJin.etal2017}, and we use the same RLS reconstruction instead of FBP to train the U-Net models. DPIR refers to an iterative deterministic method that uses deep Gaussian denoiser as prior for solving various imaging inverse problems. The denoisers used in DPIR are retrained on our CT datasets. ILVR and DPS are two sampling algorithms that use unconditionally trained diffusion models for solving inverse problems. It is worth noting that to the best of our knowledge there is no existing work that uses diffusion models for LACT reconstruction. We run a grid search over the noise schedule and data-consistency hyper-parameters for both ILVR and DPS, and we observe that both ILVR and DPS perform better in terms of PSNR/SSIM when using models trained separately on each dataset. Accordingly, we report the results that have the best PSNR (dB) values. From Table~\ref{table:results1} and Table~\ref{table:results2}, it is evident that DOLCE is significantly better than existing approaches and significantly outperforms recent methods using unconditionally trained diffusion models.

\begin{table}[t!]
  	\centering
  	\resizebox{\columnwidth}{!}
  	{\begin{tabular}{lcccccc}
	\toprule
 	\multicolumn{1}{l}{\textbf{Metric}} & \multicolumn{3}{c}{\bf PSNR $\,\uparrow$} & \multicolumn{3}{c}{\bf SSIM $\,\uparrow$}\\
  	\cmidrule{2-4} \cmidrule{5-7} 
\textbf{Angle}           & \bf 60$^\circ$ & \bf 90$^\circ$ & \bf 120$^\circ$  & \bf 60$^\circ$ & \bf 90$^\circ$ & \bf 120$^\circ$\\ 	
 	\hline\hline
FBP                                      & 25.70 & 27.87 & 31.75  & 0.673 & 0.694 & 0.739\\
RLS                                      & 27.45 & 30.69 & 34.91  & 0.756 & 0.852 & 0.909\\
TV~\cite{Beck.Teboulle2009}                & 29.13 & 33.01 & 39.06  & 0.811 & 0.902 & 0.963\\
CTNet~\cite{Anirudh.etal2018}                & 29.72 & 33.39  & 37.95  & 0.824 & 0.895 & 0.952\\
U-Net~\cite{DJin.etal2017}             & 29.47 & 33.45 & 39.22  & 0.851 & 0.910 & 0.972\\
DPIR~\cite{Zhang.etal2021b}                & 30.40 & 34.35 & 38.92  & 0.845 & 0.916 & 0.970\\
ILVR~\cite{Choi.etal2021}                 & 29.64 & 33.06 & 38.97  & 0.846 & 0.911 & 0.968\\
DPS~\cite{Chung.etal2022a}               & 30.96 & 34.84 & 38.75  & 0.885 & 0.923 & 0.968\\
\hline
DOLCE                                    & \colorbox{lightblue}{\makebox(24,4){34.06}}  & \colorbox{lightblue}{\makebox(24,4){39.01}}  & \colorbox{lightblue}{\makebox(24,4){44.83}}  & \colorbox{lightblue}{\makebox(24,4){0.932}} & \colorbox{lightblue}{\makebox(24,4){0.964}} & \colorbox{lightblue}{\makebox(24,4){0.985}}\\
DOLCE-SA                                 &  \colorbox{lightgreen}{\makebox(24,4){\bf 34.74}} & \colorbox{lightgreen}{\makebox(24,4){\bf39.67}} & \colorbox{lightgreen}{\makebox(24,4){\bf45.52}}  & \colorbox{lightgreen}{\makebox(24,4){\bf0.937}} & \colorbox{lightgreen}{\makebox(24,4){\bf0.972}} & \colorbox{lightgreen}{\makebox(24,4){\bf0.987}}\\
    \hline
	\end{tabular}}
	\caption{
	Average PSNR and SSIM results comparing test slices with the ground truth from checked-in luggage dataset.}
	\label{table:results2}
	\vspace{-1em}
\end{table}


\noindent
\textbf{Visual Evaluation.} We compare the visual results of DOLCE to RLS, TV, ILVR, and DPS for $\theta_{\text{max}}\in\{60^{\circ}, 90^{\circ}\}$ in Fig.~\ref{Fig:fig3}. In general, we observe that RLS is dominated by the artifacts due to missing angles, while TV reduces those artifacts, but blurs the fine structures by producing cartoon-like features. Although ILVR and DPS show better reconstruction with sharper edges than TV, DOLCE produces more accurate reconstructions with fine details. This highlights the SOTA performance of DOLCE using our conditionally trained denoising diffusion model.
\begin{figure}[t!]
	\centering
	\includegraphics[width=0.98\linewidth]{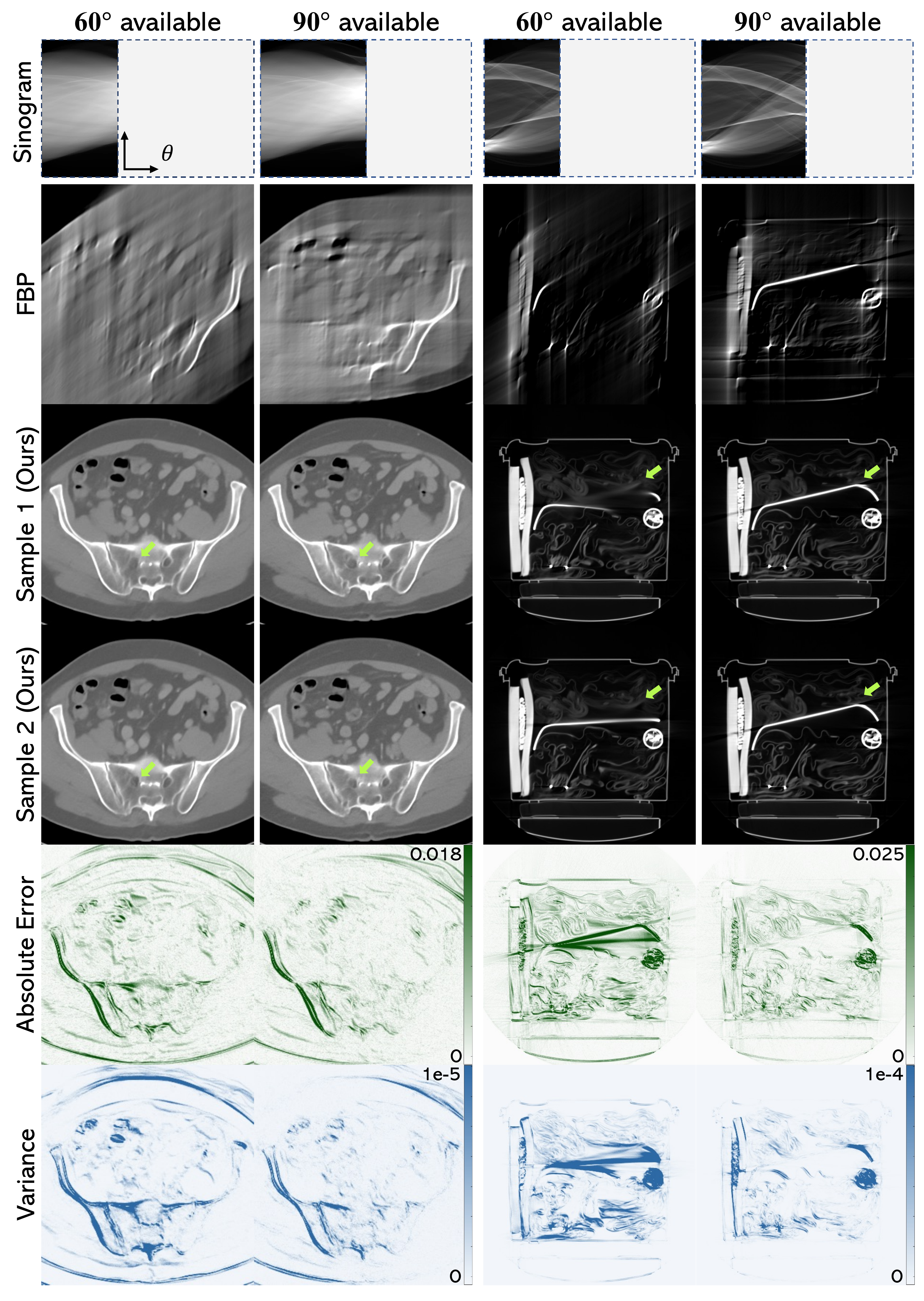}
	\caption{
    Visual results on two different CT images. The error to the ground truth is computed using the conditional mean $\E[\xbm|\ybm]$, and the variance corresponds to  per-pixel standard deviation.  It is evident that the ill-posed nature of the reconstruction task has a direct impact on the diversity of the generated samples, and the variances are highly correlated with the reconstruction errors.       
	}
 
	\label{Fig:fig4}
\end{figure}
\begin{table}[t!]
  	\centering
  	\resizebox{0.95\columnwidth}{!}
  	{\begin{tabular}{lcccc}
	\toprule
\textbf{Angle}     &  \bf Dataset    &Lug. &Med. & Lug.+Med.\\ 	
 	\hline\hline
\multirow{2}{*}{\bf60$^{\circ}$} &Lug.  & 33.59 / \textbf{0.935} & 26.78 / 0.701 & \textbf{33.98} / \textbf{0.935}\\
&Med. & 22.56 / 0.726 & 34.95 / \bf{0.949} & \textbf{35.15} / 0.945\\
\hline
\multirow{2}{*}{\bf90$^{\circ}$} &Lug.  & 39.19 / 0.966 & 31.36 / 0.853 & \bf 39.28 / 0.967\\
&Med. & 29.96 / 0.732 & \bf 39.28 / 0.969 & 39.27 / 0.963\\
\hline
\multirow{2}{*}{\bf120$^{\circ}$} &Lug.  & \bf 45.43 / 0.988 & 34.71 / 0.933 & 45.18 / 0.987\\
&Med. & 33.97 / 0.927  & \bf 43.05 / 0.976 & 42.52 / 0.974\\
    \hline
	\end{tabular}}
	\caption{
	Average PSNR/SSIM results of DOLCE on luggage and medical images, where DOLCE uses two models separately trained on luggage and medical and one trained on the combined dataset.}
	\label{table:results3}
	\vspace{-0.5em}
\end{table}

\subsection{Ablation Studies}
\noindent
\textbf{Capacity for Multiple Data Distributions.} We extract additional 150 slices randomly selected from luggage and body CT datasets, respectively, in order to evaluate the effectiveness of our DOLCE using model jointly trained on two distinct datasets (denoted as ~\emph{``Lug.+Med."}) versus models trained separately. The average PSNR/SSIM values for different limited angles are presented in Table~\ref{table:results3}. We find that DOLCE is remarkably consistent in matching the performance of the individually trained models across both domains, which highlights the potential of using a single diffusion-based CT reconstruction model to work effectively across a variety of applications. 
\begin{figure}[t!]
	\centering
	\includegraphics[width=0.95\linewidth]{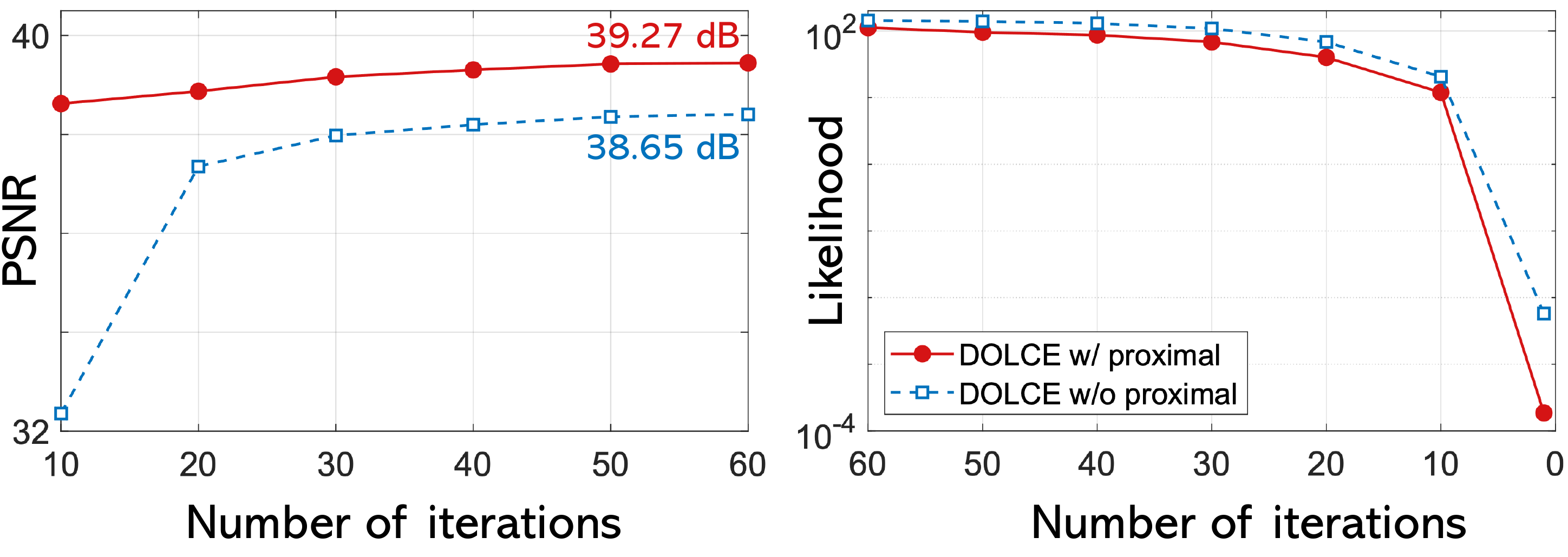}
	\caption{
    Comparison of average PSNR (left) and likelihood (right) of DOLCE w/ and w/o data-consistency mapping in Eq.~\eqref{eq:proxrule} on medical CT dataset with  $\theta_{\textbf{max}}=$90$^{\circ}$. Both methods use re-scheduling strategy of IDDPM~\cite{Nichol.etal2021} starting from $K=10$. The likelihood is plotted using $K=60$. Note the improved reconstruction quality by imposing data-consistency during inference.
	}
	\label{Fig:fig5}
	\vspace{-0.8em}
\end{figure}

\begin{figure}[b!]
	\centering
	\includegraphics[width=0.95\linewidth]{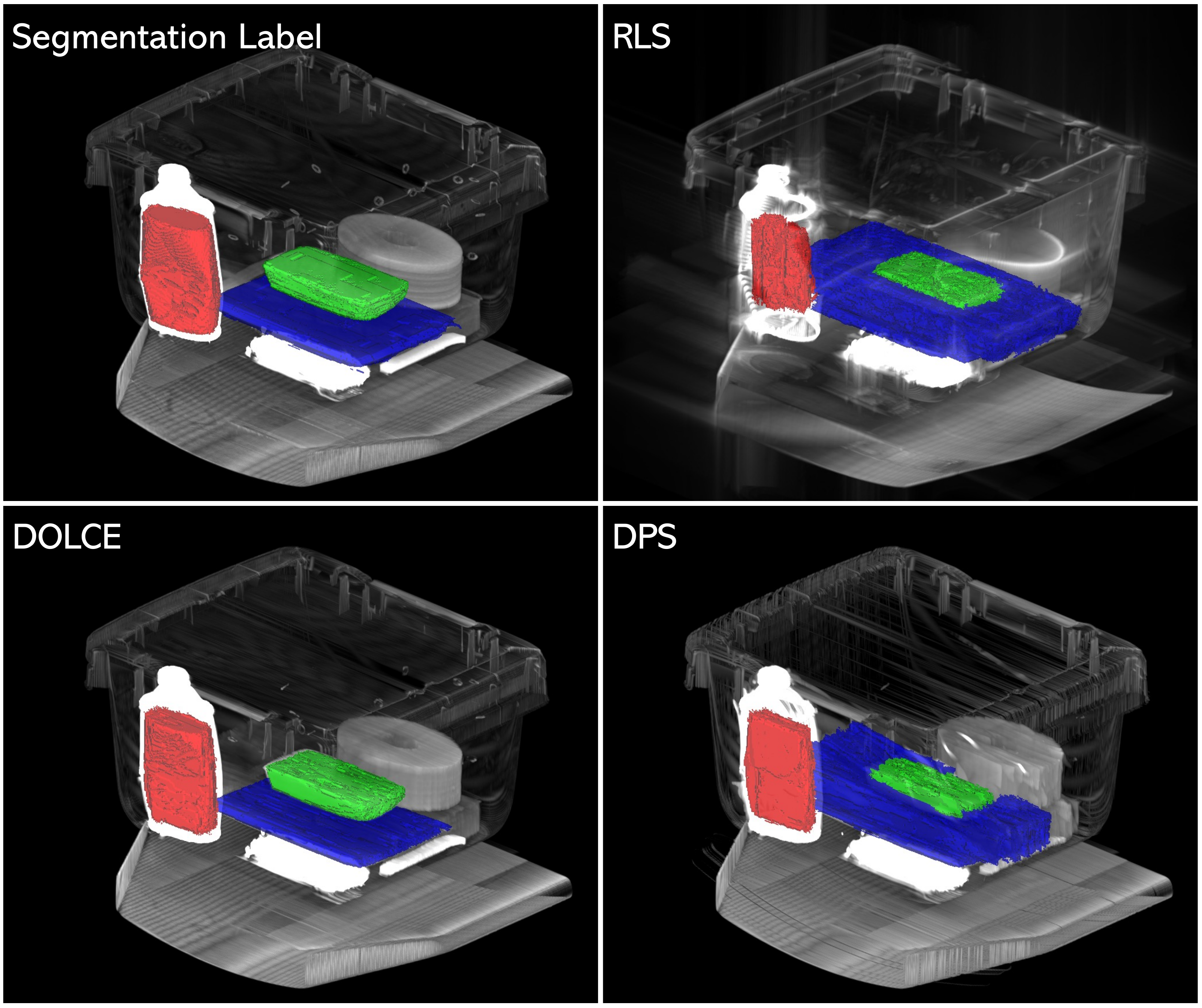}
	\caption{
	We use a region growing 3D segmentation in all cases and the resulting segmentations are highlighted in color, against a 3D rendering of the reconstructed 2D slices using $\theta_{\text{max}}=$90$^{\circ}$. Note that our method performs very similar to ground truth in determining the object boundaries compared to RLS and DPS.
	}
	\label{Fig:fig6}
\end{figure}

\noindent
\textbf{Uncertainty Quantification.} Fig.~\ref{Fig:fig4} shows that DOLCE is able to quantify uncertainty by estimating the variances directly. Since a well-calibrated model indicates larger variance in areas of larger absolute error, variance can be used as a proxy for reconstruction error in the absence of ground truth. It is evident in Fig~\ref{Fig:fig4}, that the variance images are highly correlated to the absolute error images, reflecting higher uncertainty in the corresponding regions. As expected, we also observe that the level of detail produced by our method is adaptive to the ill-posed nature of the reconstruction task, since more ill-posed input generally leads to higher variance in the resulting samples.

\medskip\noindent
\textbf{Incorporation of Data-Consistency.} Visualizing the trend of PSNR in Fig.~\ref{Fig:fig5} (left), we see that the quality of the image improves as we use more number of iterations and remains steady after $K=50$. More importantly, DOLCE using the data-consistency provided in Eq.~\eqref{eq:proxrule}~\emph{boosts} the reconstruction quality with less sampling steps. Additionally, both DOLCE w/ and w/o proximal mapping are reducing the likelihood during inference as illustrated in Fig.~\ref{Fig:fig5} (right), whereas enforcing proximal mapping leads to a lower likelihood as expected, which highlights the potential of enforcing data-consistency within sampling.

\subsection{3D Segmentation from CT Reconstructions}
Since CT images are primarily used to study 3D objects, we evaluate the quality of the DOLCE reconstructions in 3D segmentation to demonstrate its usefulness in practice. To this end, we use the popular region-growing based segmentation proposed in~\cite{Wiley.etal2012} to identify high intensity objects in the bags from their reconstructions with limited angular range. We show in Fig.~\ref{Fig:fig6} an example of a bag (from the test set) with 274 image slices that has been rendered in 3D using the 2D slices reconstructed with the proposed DOLCE. We compare the segmentations obtained using our method to the segmentation labels as reference, and those obtained using RLS and DPS, respectively. Specifically, both RLS and DPS preserves 3D edges poorly resulting in spurious segments, whereas our DOLCE reconstruction is significantly better, resembling the ground truth. Additional segmentation results can be found in the supplementary material.

\section{Conclusion}
We consider the recovery of high-quality images from the LACT data in the settings where the viewing angles can be as small as 60$^{\circ}$. Building on the recent work on conditional diffusion models, we present the first model-based probabilistic diffusion framework for LACT called DOLCE. Our framework enables the recovery of high-quality CT images that preserve the geometric structure and sharp edges by using an image prior in the form of a diffusion model conditioned on the transformed limited-angle sinograms. DOLCE can use FBP or RLS images as the conditional input to its diffusion model. During inference, DOLCE enforces the forward model using the data-consistency update implemented as a proximal mapping. As a result, DOLCE imposes both forward-model and prior constraints on the solution. Extensive experimental results demonstrate the SOTA performance of DOLCE on widely different data distributions, such as images of human body and of checked-in luggage, thus enabling highly generalizable LACT reconstruction networks for the first time. Additionally, we show how the diverse realizations produced by DOLCE from a given sinogram can enable meaningful uncertainty quantification. In summary, our work presents a new SOTA method for LACT that enables systematic uncertainty characterization, thus opening a new exciting avenue for future research on diffusion models for severely ill-posed imaging problems such as LACT.

\section*{Acknowledgment}

\noindent
This work was performed under the auspices of the U.S. Department of Energy by Lawrence Livermore National Laboratory under Contract DE-AC52-07NA27344. This work was funded by the Laboratory Directed Research and Development (LDRD) program at Lawrence Livermore National Laboratory (22-ERD-032). LLNL-CONF-816780. This material is based upon work supported by the U.S. Department of Homeland Security, Science and Technology Directorate, Office of University Programs, under Grant Award 2013-ST-061-ED0001. The views and conclusions contained in this document are those of the authors and should not be interpreted as necessarily representing the official policies, either expressed or implied, of the U.S. Department of Homeland Security. 


\newpage 

\appendix

{\Large\textbf{~~~~~Supplementary Material}}
\vspace{1em}
\medskip

\section{CT Reconstruction Formulation}
In our experiments, the object to be imaged is placed in between a source of parallel
beam x-rays and a planar detector array. The x-rays get attenuated as they propagate through the object and the intensity of attenuated x-rays exiting the object is measured by the detector. To perform tomographic imaging, the object is rotated along an axis and repeatedly imaged at regular angular intervals of rotation. Assume that the object is stationary in the Cartesian coordinate system represented by the axes $(x,y,z)$, at each rotation angle $\theta$ of the object, we are interested in reconstructing 2D slice images, denoted as $\rho(x,y,z)$ of object linear attenuation coefficient (LAC) values along the propagation path. The projection at a distance of $r$ on the detector is given by
\begin{equation}
\begin{aligned}
    \label{eq:radon}
    &S_{\theta}(r,z) = \int\int  \rho(x,y,z)\delta(x\cos(\theta) + y\sin(\theta) -r) \,dxdy, \\
\end{aligned}    
\end{equation}
where $\delta$ is the indicator function and $S_{\theta}(r,z)$ is known as sinogram. Note that equation~\eqref{eq:radon} is separable in the $z$ coordinate. Hence, the projection relation is essentially a 2D function in the $x-y$ plane that is repeatedly applied along the $z$-axis. The reconstruction of $\rho(x,y,z)$ from incomplete sinogram can be formulated as image inverse problem described in the main paper (See~\cite{Kak.etal2001, Radon.etal1986} for more references). 

\medskip\noindent
\textbf{FBP Reconstruction}
Filtered back-projection (FBP) is an analytic algorithm for reconstructing the sample $\rho(x,y,z)$ for the projections $S_{\theta}(r,z)$ at all the rotation angles $\theta$. In FBP, we first compute the filtered projection measurement of each slice $\widehat S_{\theta}(r) = \int \Fcal[S_{\theta}](\omega)|\omega|e^{j2\pi\omega r}d\omega$, where $\Fcal$ denotes the Fourier transform and $|\omega|$ is the frequency response of the filter. The filtered back projection reconstruction is then given by~\cite{Kak.etal2001} 
\begin{equation}
\begin{aligned}
    \label{eq:fbp}
    &f_{\text{FBP}}(x,y) = \int_{0}^{\pi}\widehat S_{\theta}(x\cos(\theta)+y\sin(\theta)) d\theta. \\
\end{aligned}    
\end{equation}
According to equation~\eqref{eq:fbp}, we know that a filtered version of $S_{\theta}$ is smeared back on the $x-y$ plane along the direction ($\pi/2-\theta$). The FBP reconstruction thus consists of the cumulative sum of the smeared contributions from all the projections ranging in $0<\theta<\pi$.

\begin{figure}[t!]
	\centering
	\includegraphics[width=0.99\linewidth]{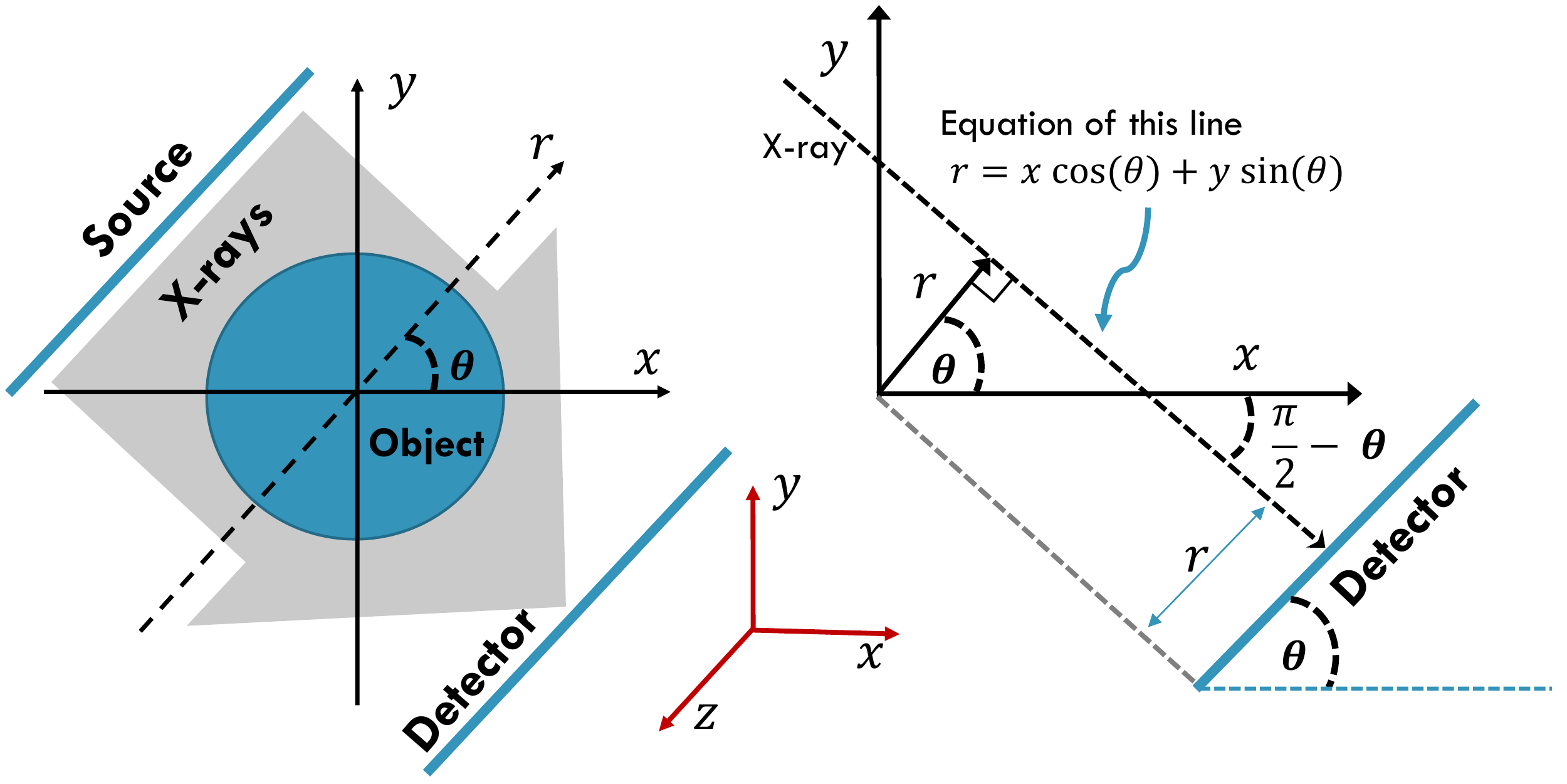}
	\caption{
Implementation of x-ray CT. An object is rotated along an axis and exposed to a parallel beam of x-rays. The intensity of attenuated x-rays exiting the object is measured by the detector at regular angular intervals. The projection at an angle of $\theta$ measured at a distance of $r$ on the detector is the line integral of LAC values along the line perpendicular to the detector.
    }
    \label{Fig:fig1_sup}
\end{figure}

\medskip\noindent
\textbf{LACT Reconstruction Artifacts by FBP.} If the projections are acquired over a limited angular range, then the integration in~\eqref{eq:fbp} will be incomplete in the angular space. Since each projection $S_{\theta}(r)$ contains the cumulative sum of the LAC values at a rotation angle of $\theta$, it also contains information about the edges that are oriented along the angular direction ($\pi/2-\theta$) as shown in Fig.~\ref{Fig:fig1_sup}. Now, suppose data acquisition starts at $\theta=\text{0}$ and stops at an angle of $\theta = \theta_{\text{max}} < \pi$. Then, the edge information contained in the projections at the angles $\theta \in [\theta_{\text{max}}, \pi]$ will be missing in the final reconstruction. This is the reason behind the edge blur in the FBP reconstructions shown in this paper.
\section{Additional Implementation Details} 
\begin{table*}[t!]
  	\centering
  	\resizebox{0.95\linewidth}{!}
  	{\begin{tabular}{lm{20pt}m{40pt}m{40pt}m{40pt}m{40pt}m{40pt}m{40pt}}
	\toprule
 	\multicolumn{1}{l}{\textbf{Metric}} & & \multicolumn{3}{c}{\bf PSNR $\,\uparrow$} & \multicolumn{3}{c}{\bf SSIM $\,\uparrow$}\\
  	\cmidrule{3-5} \cmidrule{6-8} 
\textbf{Angle}       &  & \textbf{$\;\;$60}$^\circ$ & \textbf{$\;\;$90}$^\circ$ & \textbf{$\;\;$120}$^\circ$  & \textbf{$\;\;$60}$^\circ$ & \textbf{$\;\;$90}$^\circ$ & \textbf{$\;\;$120}$^\circ$\\ 	
 	\hline\hline
FBP                                      && 14.95 & 17.28 & 20.97  & 0.464 & 0.543 & 0.603\\
RLS                                      && 22.72 & 26.19 & 30.42  & 0.699 & 0.833 & 0.888\\
DOLCE (FBP, w/o prox)                && 33.72 & 38.00 & 40.63  & 0.927 & 0.945 & 0.954\\
DOLCE (FBP, w/ prox)                && 34.05 & 38.73 & 42.00  & 0.938 & 0.960 & 0.972\\
DOLCE (RLS, w/o prox)                && 34.91 & 38.65 & 41.25  & 0.936 & 0.951 & 0.959\\
DOLCE (RLS, w/ prox)                && 35.15 & 39.27 & 42.52  & 0.945 & 0.963 & 0.974\\
\hline
DOLCE-SA (FBP, w/o prox)                && 34.31 & 38.75 & 41.49  & 0.936 & 0.952 & 0.961\\
DOLCE-SA (FBP, w/ prox)                && 34.58 & 39.28 & 42.55  & 0.943 & 0.963 & 0.975\\
DOLCE-SA (RLS, w/o prox)                && 35.55 & 39.31 & 42.12  & 0.944 & 0.954 & 0.965\\
DOLCE-SA (RLS, w/ prox)                && \bf35.78 & \bf39.75 & \bf43.11  & \bf0.949 & \bf0.969 &\bf0.977
\\
    \hline
	\end{tabular}}
	\caption{
	Average PSNR and SSIM results comparing test slices with the ground truth from medical body CT dataset.}
	\label{table:results3_sup}
	\vspace{1em}
\end{table*}

\begin{table*}[t!]
  	\centering
  	\resizebox{0.95\linewidth}{!}
  	{\begin{tabular}{lm{20pt}m{40pt}m{40pt}m{40pt}m{40pt}m{40pt}m{40pt}}
	\toprule
 	\multicolumn{1}{l}{\textbf{Metric}} & & \multicolumn{3}{c}{\bf PSNR $\,\uparrow$} & \multicolumn{3}{c}{\bf SSIM $\,\uparrow$}\\
  	\cmidrule{3-5} \cmidrule{6-8} 
\textbf{Angle}       & ~~ & \textbf{$\;\;$60}$^\circ$ & \textbf{$\;\;$90}$^\circ$ & \textbf{$\;\;$120}$^\circ$  & \textbf{$\;\;$60}$^\circ$ & \textbf{$\;\;$90}$^\circ$ & \textbf{$\;\;$120}$^\circ$\\ 	
 	\hline\hline
FBP                                     &&26.08&28.34&32.18& 0.668& 0.713&0.752\\
RLS                                      && 28.05 & 31.01 & 35.61  & 0.775 & 0.860 & 0.914\\
DOLCE (FBP, w/o prox)                && 33.13 & 37.43 & 42.83  & 0.928 & 0.953 & 0.974\\
DOLCE (FBP, w/ prox)                && 33.44 & 38.17 & 44.12  & 0.928 & 0.959 & 0.983\\
DOLCE (RLS, w/o prox)                && 33.70 & 38.64 & 43.76  & 0.933 & 0.959 & 0.978\\
DOLCE (RLS, w/ prox)                && 33.98 & 39.28 & 45.18  & 0.935 & 0.967 & 0.987\\
\hline
DOLCE-SA (FBP, w/o prox)                && 33.85 & 38.26 & 43.74  & 0.932 & 0.959 & 0.978\\
DOLCE-SA (FBP, w/ prox)                && 34.11 & 38.91 & 44.87  & 0.931 & 0.963 & 0.985\\
DOLCE-SA (RLS, w/o prox)                &&  34.41 & 39.40 & 44.52  & \bf0.941 & 0.966 & 0.981\\
DOLCE-SA (RLS, w/ prox)                && \bf34.68 & \bf39.88 & \bf45.68  & \bf0.941 & \bf0.971 & \bf0.988
\\
    \hline
	\end{tabular}}
	\caption{
	Average PSNR and SSIM results comparing test slices with the ground truth from checked-in luggage dataset.}
	\label{table:results4_sup}
    \vspace{1em}
\end{table*}

\noindent
\textbf{CTNet}~\cite{Anirudh.etal2018} is an end-to-end DL method, designed to predict the invisible sinogram data by incorporating a GAN into the neural network architecture.  Note that the original CTNet was developed on $128\times128$ images. To make it work on $512\times512$ images, given the pre-trained CTNet on $128\times128$ images, we additionally train a super-resolution diffusion model presented in~\cite{Saharia.etal2022a} to super-resolve the low-resolution outputs of CTNet to the same $512\times512$ dimension as other methods.

\medskip\noindent
\textbf{DPIR}~\cite{Zhang.etal2021b} refers to the SOTA PnP methods using deep denoiser as prior for solving various ill-posed image inverse problems. We modify the publicly available implementation to train the deep denoiser~\footnote{\url{https://github.com/cszn/KAIR}} on each dataset separately and follow the similar implementation settings~\footnote{\url{https://github.com/cszn/DPIR}} at inference. Since our CT images are naturally in smaller intensity range, we train the DPIR denoiser for the AWGN removal within noise level of $\sigma\in[0,5]$.

\medskip\noindent
\textbf{ILVR}~\cite{Choi.etal2021} and \textbf{DPS}~\cite{Chung.etal2022a} refer to recently developed conditioning methods based on unconditionally trained DDPM for solving versatile ill-posed inverse problems. We modify the publicly available implementation of ILVR~\footnote{\url{https://github.com/jychoi118/ilvr_adm}} and DPS~\footnote{\url{https://github.com/DPS2022/diffusion-posterior-sampling}} in order to incorporate our LACT forward-model. We use the similar grid search as DOLCE for fine-tuning the hyper-parameters within ILVR and DPS, respectively.

\begin{figure*}[t!]
	\centering
	\includegraphics[width=0.95\textwidth]{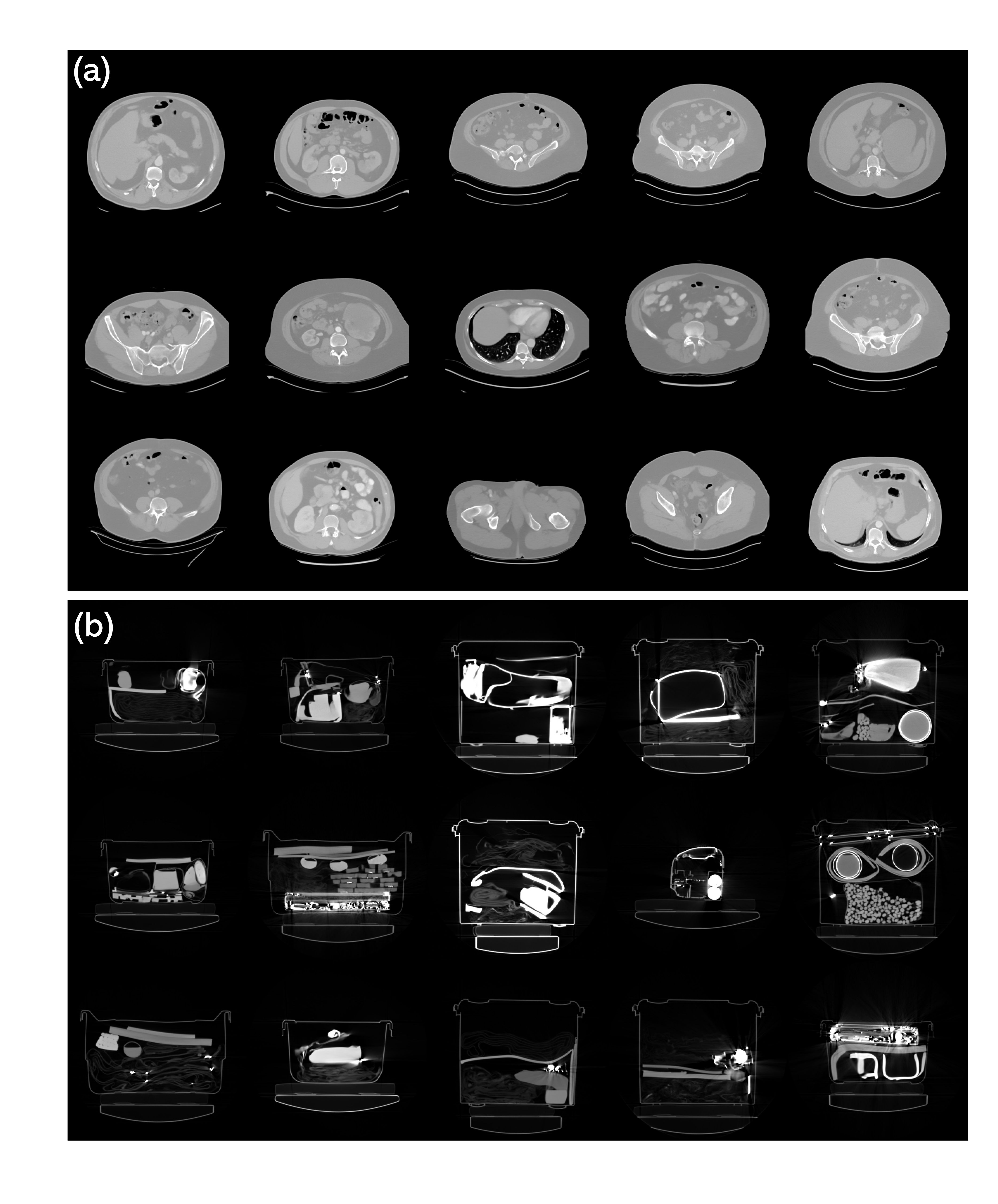}
	\caption{Random samples from our two unconditionally trained denoising diffusion $512\times512$ models, respectively. \textbf{(a):} diffusion model trained on human body CT images; \textbf{(b):} diffusion model trained on checked-in luggage dataset. These models are used in ILVR~\cite{Choi.etal2021} and DPS~\cite{Chung.etal2022a} as baseline methods. Images are normalized for better visualization.
	}
	\label{Fig:rdmsample_sup}
\end{figure*}

\medskip\noindent
We train all the diffusion models used in this paper, modified based on the publicly available PyTorch implementation~\footnote{\url{https://github.com/openai/guided-diffusion}}. To indicate the high quality of our pre-trained diffusion models used within ILVR and DPS, we present the random samples from our two unconditionally trained denoising diffusion $512\times512$ models in Fig.~\ref{Fig:rdmsample_sup} for luggage and medical dataset, respectively.

\section{Additional Numerical and Visual Results}

\begin{table*}[h!]
  	\centering
  	\resizebox{0.95\linewidth}{!}
  	{\begin{tabular}{lm{20pt}m{45pt}m{45pt}m{45pt}m{45pt}m{45pt}m{45pt}}
	\toprule
 	\multicolumn{1}{l}{\textbf{Metric}} && \multicolumn{3}{c}{\bf PSNR $\,\uparrow$} & \multicolumn{3}{c}{\bf SSIM $\,\uparrow$}\\
  	\cmidrule{3-5} \cmidrule{6-8} 
\textbf{Angle}         & ~~& \textbf{$\;\;$45}$^\circ$ & \textbf{$\;\;$50}$^\circ$ & \textbf{$\;\;$55}$^\circ$  & \textbf{$\;\;$45}$^\circ$ & \textbf{$\;\;$50}$^\circ$ & \textbf{$\;\;$55}$^\circ$\\
 	\hline\hline
FBP                                      &&14.49&14.65&14.83&0.321&0.397&0.441\\
RLS                                      &&  19.01 & 19.98 & 20.95  & 0.559 & 0.596 & 0.632\\
ILVR~\cite{Choi.etal2021}                 && 23.85 & 24.94 & 26.75  & 0.815 & 0.839 & 0.872\\
DPS~\cite{Chung.etal2022a}               && 24.50 &  25.84 & 26.71  & 0.833 & 0.848 &  0.870\\
\hline
DOLCE                                    && 24.97 & 27.47 & 31.02 & 0.838 & 0.884 & 0.934\\
DOLCE-SA                                 && \bf25.68 & \bf27.88 & \bf31.51  & \bf0.841 & \bf0.889 & \bf0.939\\
    \hline
	\end{tabular}}
	\caption{
	Average PSNR and SSIM results for several methods on body CT dataset. \textbf{Best values} for each metric are highlighted.}
	\label{table:results1_sup}
 	\vspace{1em}
\end{table*}

\begin{table*}[h!]
  	\centering
  	\resizebox{0.95\linewidth}{!}
  	{\begin{tabular}{lm{20pt}m{45pt}m{45pt}m{45pt}m{45pt}m{45pt}m{45pt}}
	\toprule
 	\multicolumn{1}{l}{\textbf{Metric}} & & \multicolumn{3}{c}{\bf PSNR $\,\uparrow$} & \multicolumn{3}{c}{\bf SSIM $\,\uparrow$}\\
  	\cmidrule{3-5} \cmidrule{6-8} 
\textbf{Angle}           & & \textbf{$\;\;$45}$^\circ$ & \textbf{$\;\;$50}$^\circ$ & \textbf{$\;\;$55}$^\circ$  & \textbf{$\;\;$45}$^\circ$ & \textbf{$\;\;$50}$^\circ$ & \textbf{$\;\;$55}$^\circ$\\ 		
 	\hline\hline
FBP                                      && 24.59 & 24.77 & 25.64  & 0.653 & 0.658 & 0.661\\
RLS                                      && 26.62 &  26.95 & 27.31  & 0.723 & 0.748 & 0.758\\
ILVR~\cite{Choi.etal2021}                 && 28.99 & 29.15 & 29.85  & 0.856 & 0.867 & 0.871\\
DPS~\cite{Chung.etal2022a}               && 29.42 & 29.85 & 30.40  & 0.862 & 0.871 & 0.878\\
\hline
DOLCE                                    && 30.46 & 31.45 & 32.39  & 0.884 & 0.899 & 0.921\\
DOLCE-SA                                 && \bf30.98 & \bf31.79 & \bf32.98  & \bf0.890 & \bf0.906 & \bf0.925\\
    \hline
	\end{tabular}}
	\caption{
	Average PSNR and SSIM results on luggage dataset.}
	\label{table:results2_sup}
	\vspace{1em}
\end{table*}
\noindent
\textbf{Comparison of RLS and FBP as Conditional Input.} In Table~\ref{table:results3_sup} and Table~\ref{table:results4_sup}, we present additional numerical evaluations for FBP and RLS as conditional input of our DOLCE models cross various angular ranges (\eg, $\theta_{\text{max}}\in\{60^{\circ},90^{\circ},120^{\circ}\}$). We use the same random selected 300 images from test luggage and medical dataset as in the main paper, respectively. While DOLCE using FBP as conditional input provides substantial improvements, using RLS input further boost the overall performance.   

\noindent
\textbf{Incorporation of Data-Consistency.} We also report additional numerical validations on the incorporation of forward-model at inference stage in Table~\ref{table:results3_sup} and Table~\ref{table:results4_sup}. We observer that DOLCE using the data-consistency provided by the proximal-mapping produces better quality reconstruction samples, which highlights the potential of enforcing forward-model within sampling step.

\medskip\noindent
\textbf{Behavior of DOLCE to Model Mismatch.} In Table~\ref{table:results1_sup} and Table~\ref{table:results2_sup}, we demonstrate the behavior of our proposed approach for varying number of views during testing. Specifically, we consider the DOLCE for forward-model mismatch scenarios, where the pre-trained DOLCE models in the main paper are tested for $\theta_{\text{max}}\in\{45^{\circ},50^{\circ},55^{\circ}\}$ limited-angle data. For references, we compare DOLCE and DOLCE-SA (sample average) to FBP, RLS, ILVR, and DPS. We can see that our DOLCE consistently outperforms baseline methods even under model mismatch cases.  

\begin{figure*}[t!]
	\centering
	\includegraphics[width=0.96\textwidth]{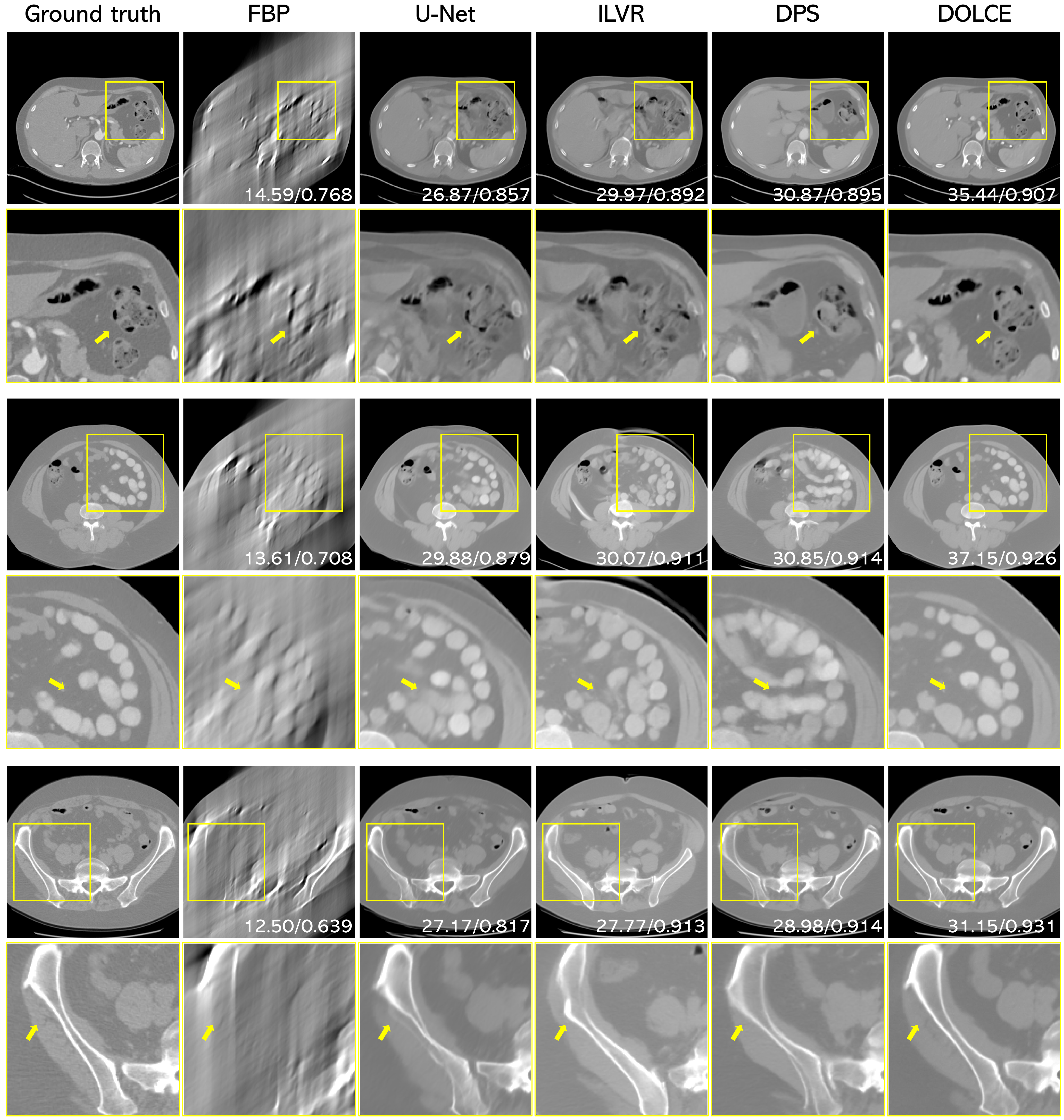}
	\caption{Visual evaluation of limited angle tomographic reconstruction in body CT images, where the input measurements are captured respectively from an angular coverage of 60$^{\circ}$. Images are normalized for better visualization.
	}
	\label{Fig:ckc1_sup}
\end{figure*}

\begin{figure*}[t!]
	\centering
	\includegraphics[width=0.96\textwidth]{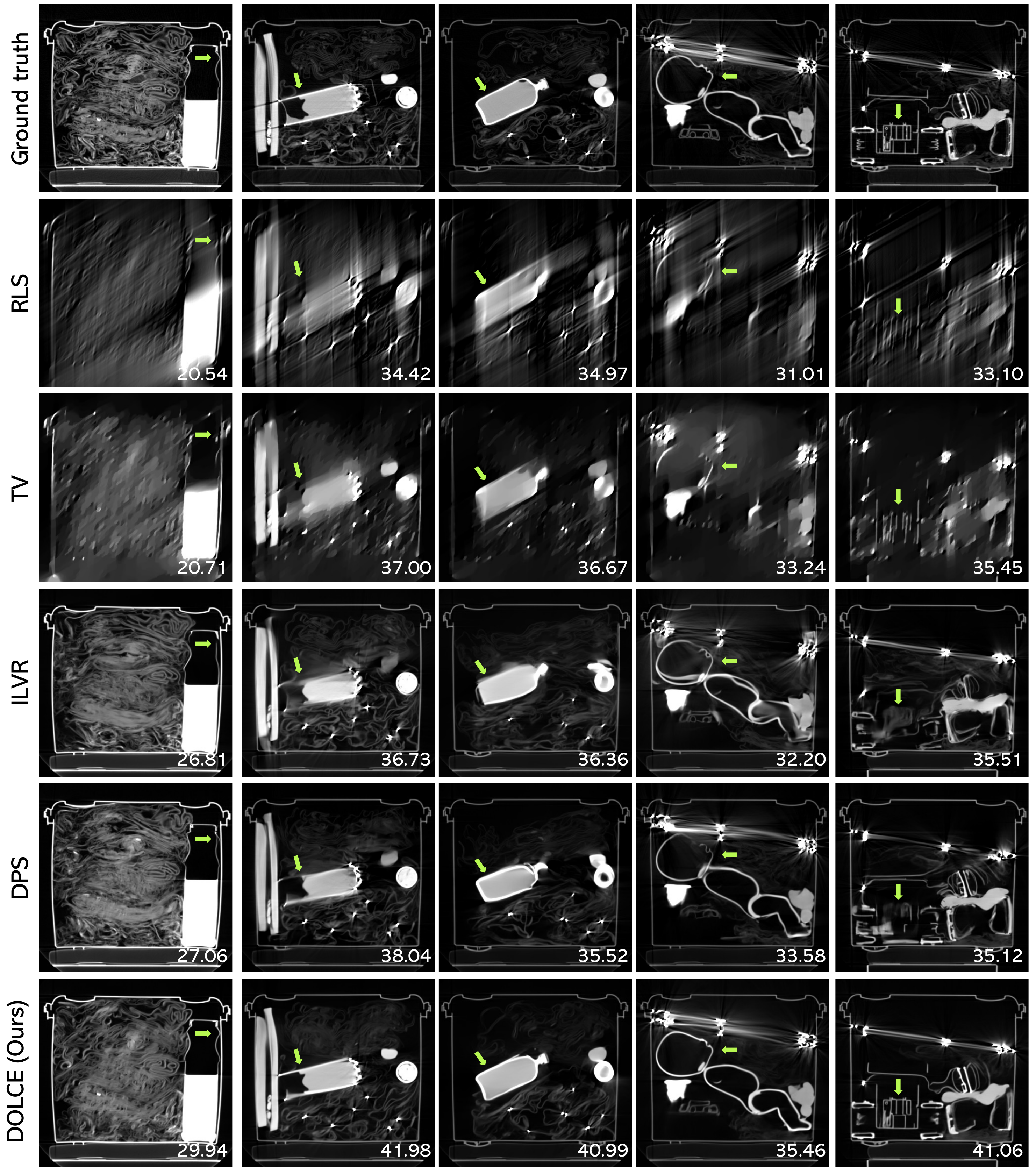}
	\caption{Visual evaluation of limited angle tomographic reconstruction in checked-in luggage, where the input measurements are captured respectively from an angular coverage of 60$^{\circ}$. Images are normalized for better visualization.
	}
	\label{Fig:Lug_Vis1_sup}
\end{figure*}

\begin{figure*}[t!]
	\centering
	\includegraphics[width=0.96\textwidth]{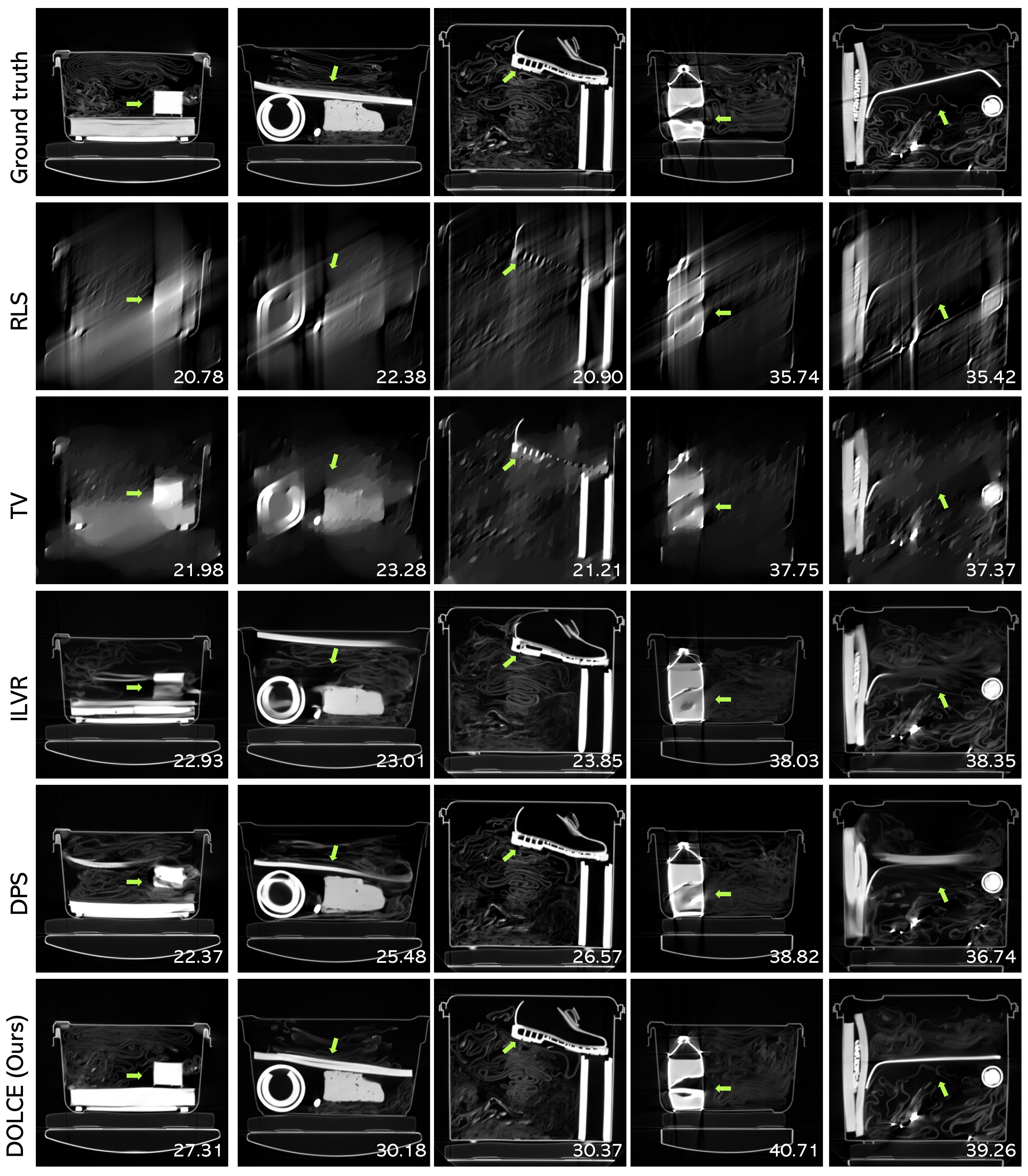}
	\caption{Visual evaluation of limited angle tomographic reconstruction in checked-in luggage, where the input measurements are captured respectively from an angular coverage of 60$^{\circ}$. Images are normalized for better visualization.
	}
	\label{Fig:Lug_Vis2_sup}
\end{figure*}

\begin{figure*}[t!]
	\centering
	\includegraphics[width=0.96\textwidth]{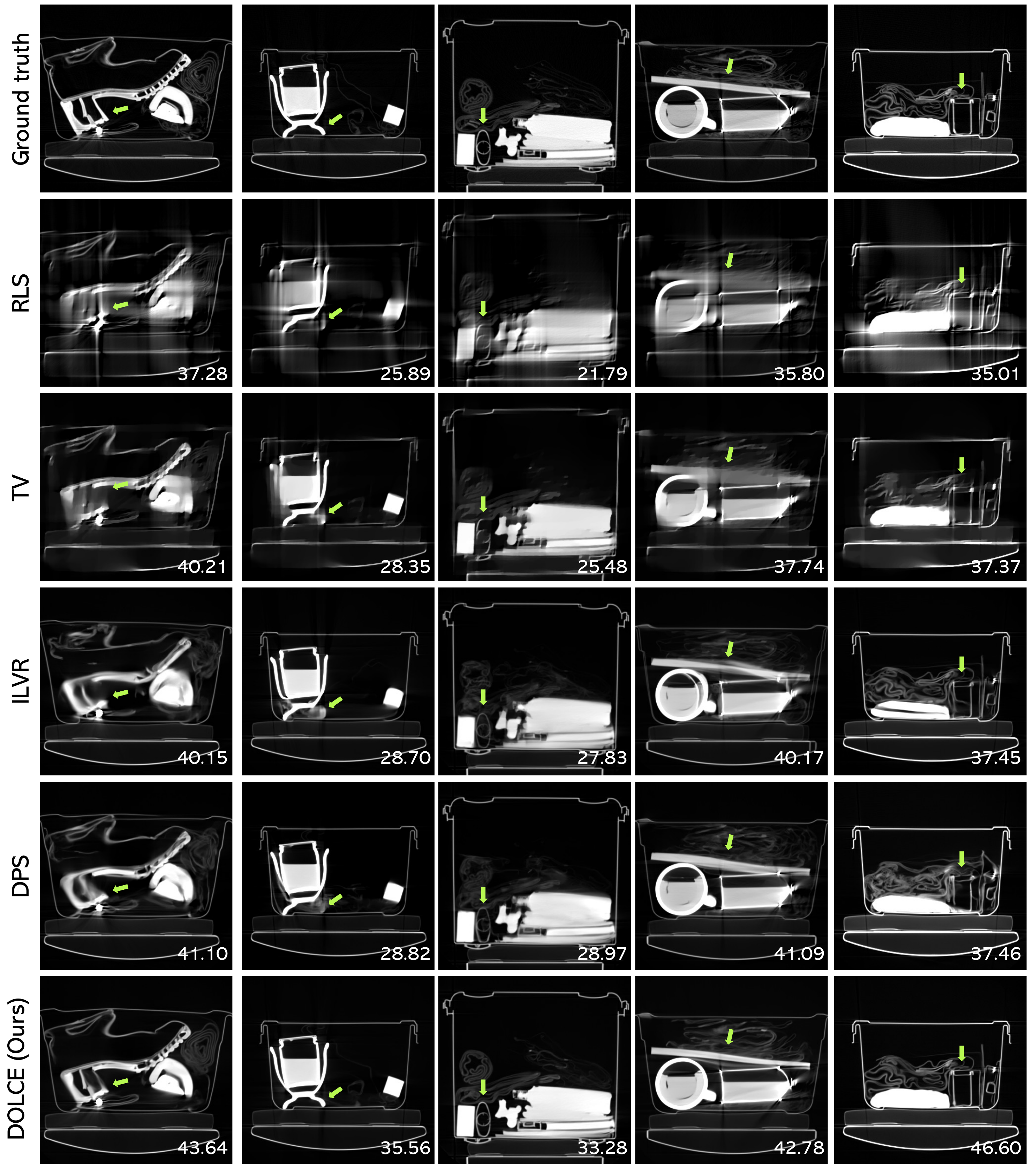}
	\caption{Visual evaluation of limited angle tomographic reconstruction in checked-in luggage, where the input measurements are captured respectively from an angular coverage of 90$^{\circ}$. Images are normalized for better visualization.
	}
	\label{Fig:Lug_Vis3_sup}
\end{figure*}

\begin{figure*}[t!]
	\centering
	\includegraphics[width=0.86\textwidth]{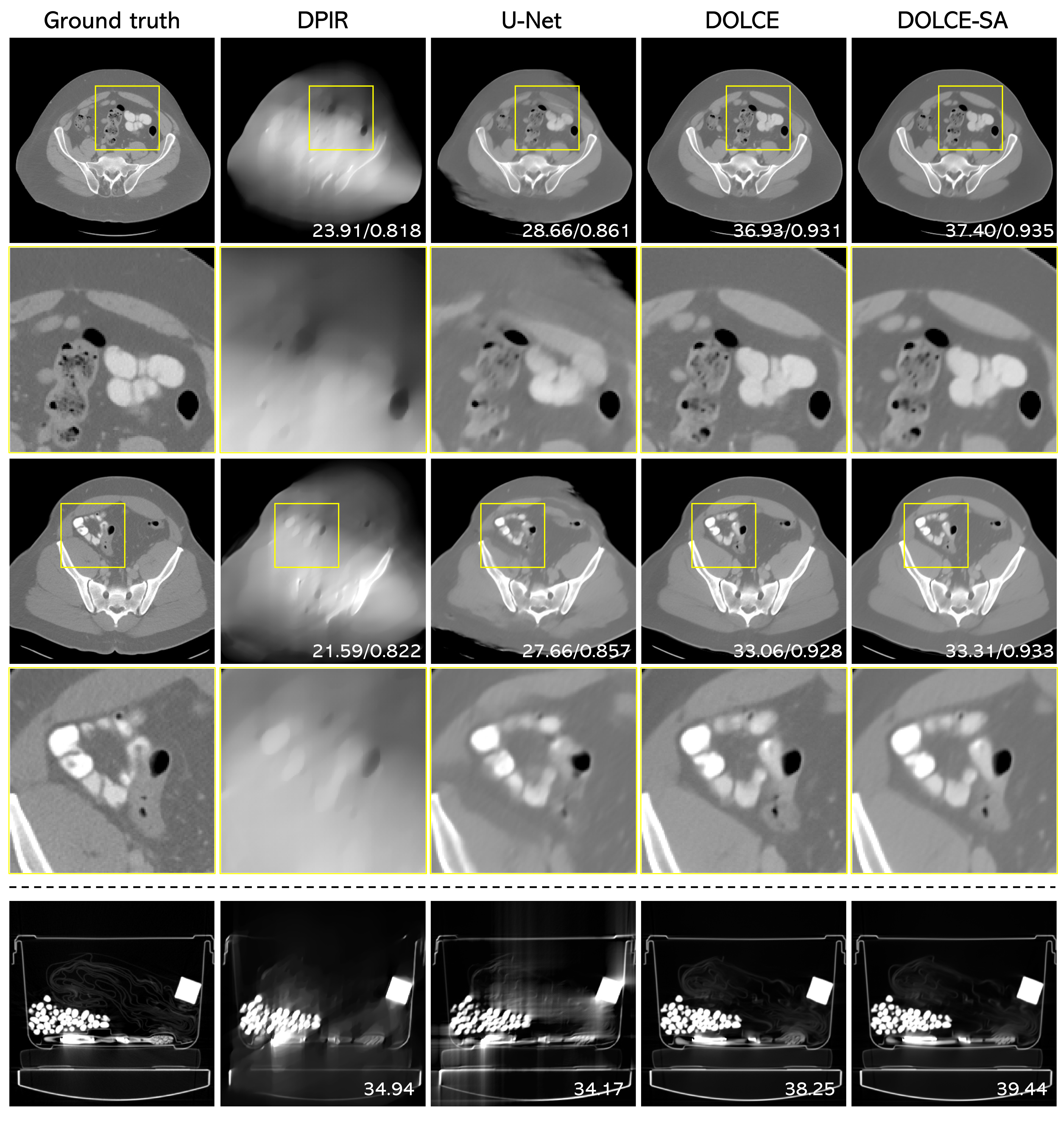}
	\caption{Additional visual evaluation of limited angle tomographic reconstruction in body CT scan (\textbf{top}) and checked-in luggage (\textbf{bottom}), where the input measurements are captured respectively from an angular coverage of 60$^{\circ}$. Images are normalized for better visualization.
	}
	\label{Fig:vis_mean_sup}
\end{figure*}

\medskip\noindent
\textbf{Additional Visual Evaluation.} In Fig.~\ref{Fig:ckc1_sup}, we compare the visual results of DOLCE on medical CT test images to FBP, U-Net, ILVR, and DPS for $\theta_{\max}=60^{\circ}$. Fig.~\ref{Fig:Lug_Vis1_sup}-\ref{Fig:Lug_Vis3_sup} present additional visual comparison for several methods on the luggage dataset. In Fig.~\ref{Fig:vis_mean_sup}, we compare DOLCE and DOLCE-SA to DPIR and U-Net on each test dataset, respectively. We also provide ~\emph{video} comparisons of our DOLCE reconstruction results in the supplement material.

\begin{figure*}[t!]
	\centering
	\includegraphics[width=0.86\textwidth]{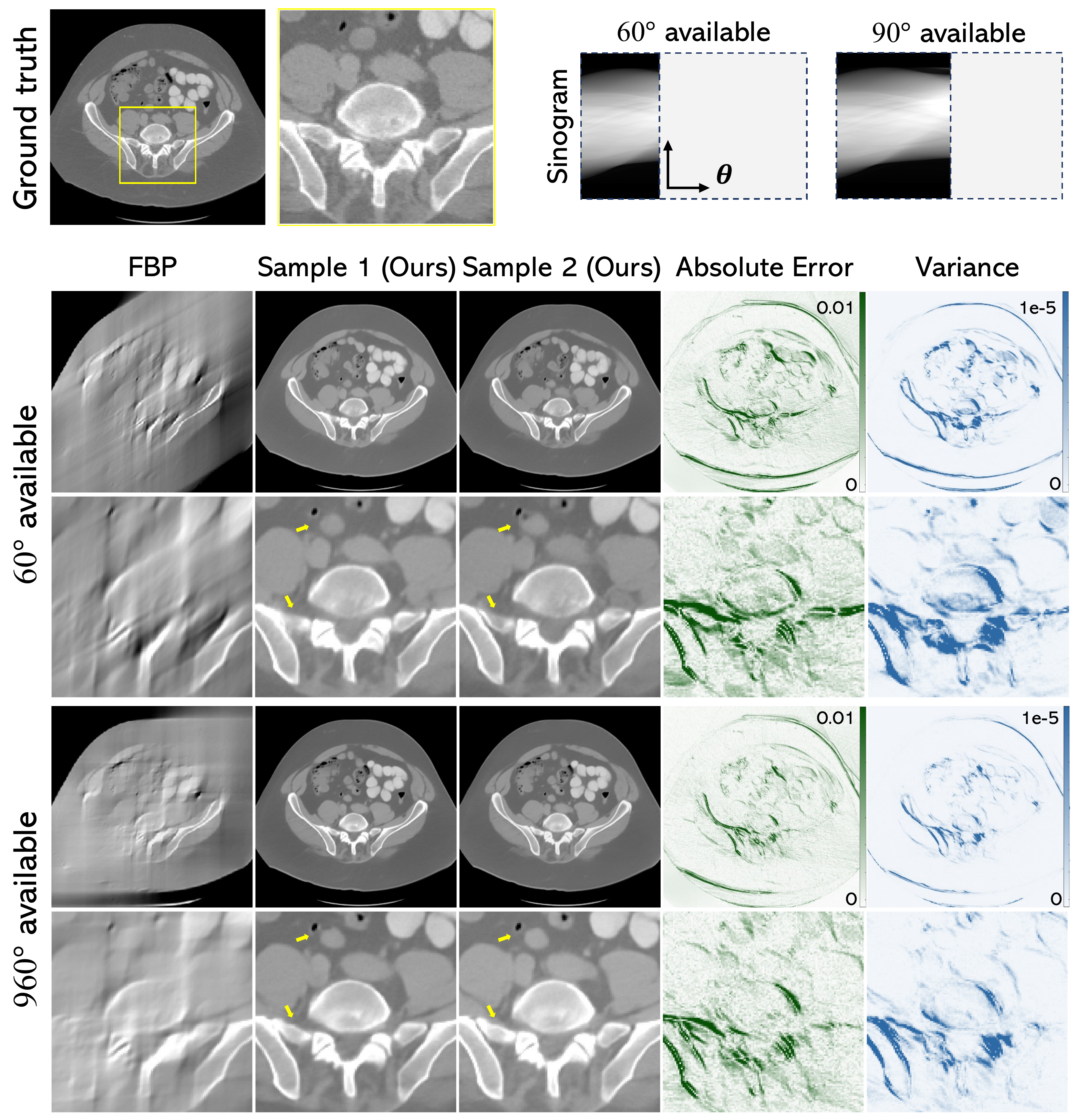}
	\caption{More visual results on body CT images. The error to the ground truth is computed using the conditional mean $\E[\xbm|\ybm]$, and the variance corresponds to  per-pixel standard deviation.  It is evident that the ill-posed nature of the reconstruction task has a direct impact on the diversity of the generated samples, and the variances are highly correlated with the reconstruction errors.
	}
	\label{Fig:ckc_mv1_sup}
\end{figure*}

\begin{figure*}[t!]
	\centering
	\includegraphics[width=0.86\textwidth]{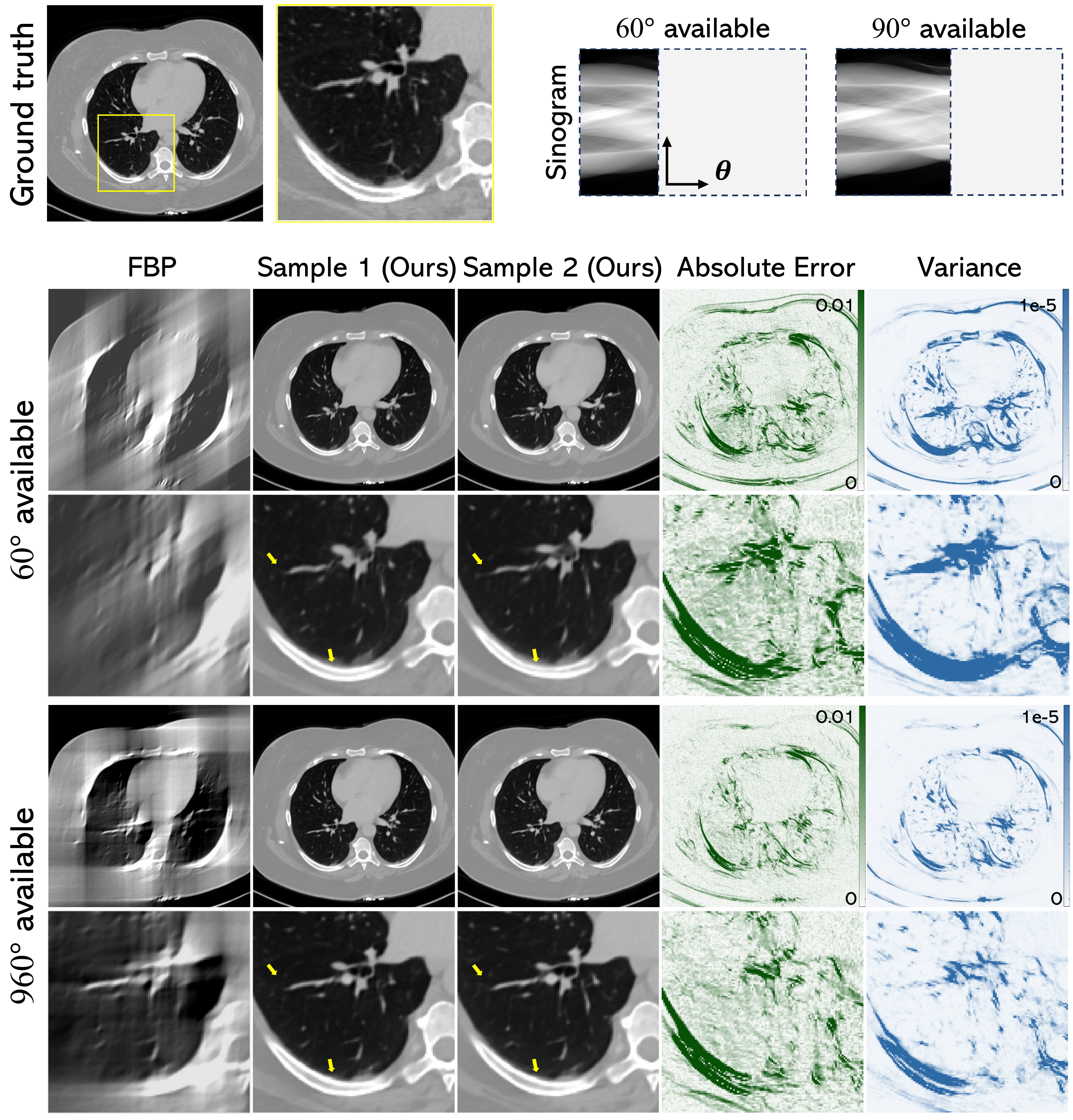}
	\caption{More visual results on body CT images. The error to the ground truth is computed using the conditional mean $\E[\xbm|\ybm]$, and the variance corresponds to  per-pixel standard deviation.  It is evident that the ill-posed nature of the reconstruction task has a direct impact on the diversity of the generated samples, and the variances are highly correlated with the reconstruction errors.  
	}
	\label{Fig:ckc_mv2_sup}
\end{figure*}

\begin{figure*}[t!]
	\centering
	\includegraphics[width=0.96\textwidth]{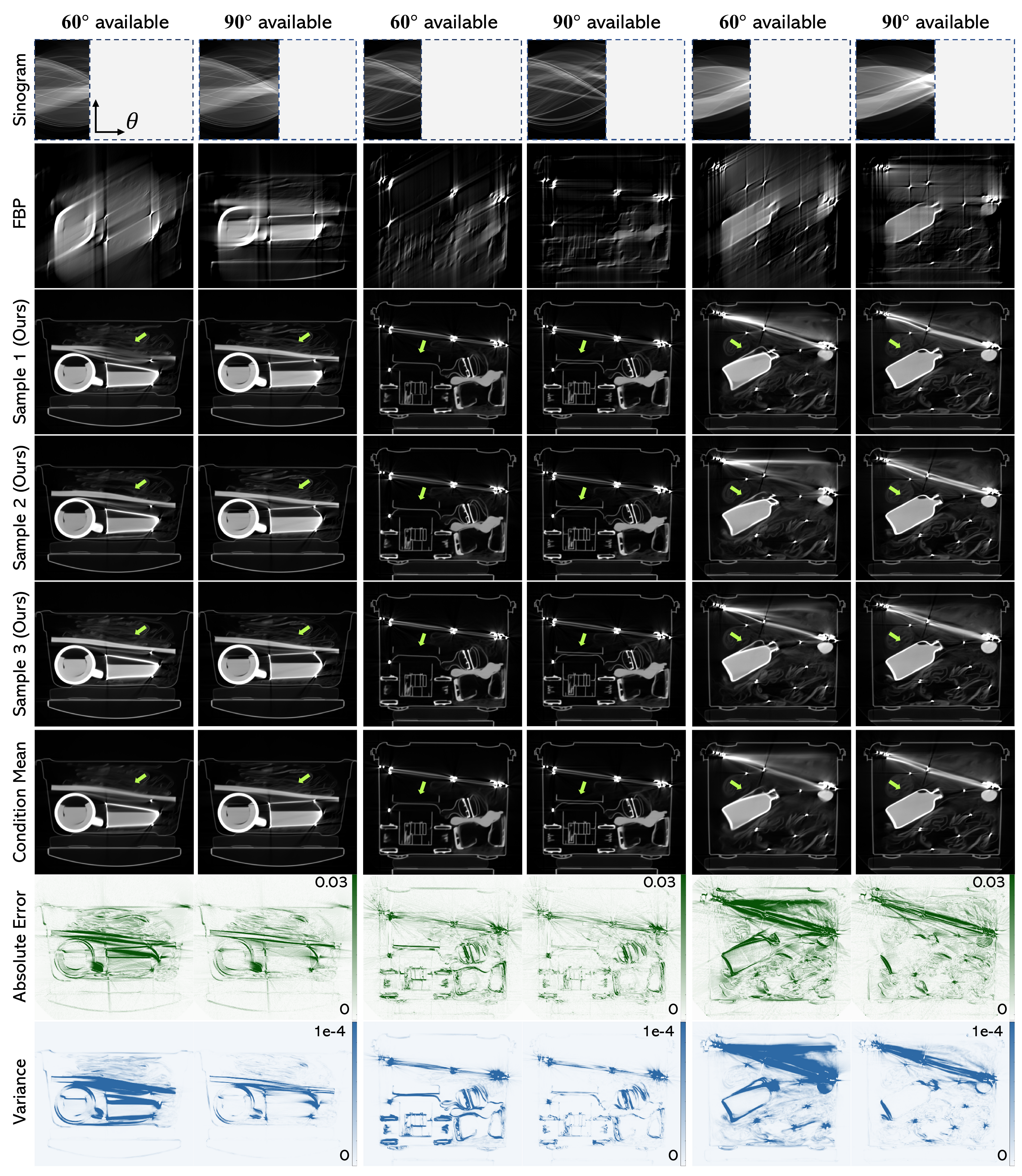}
	\caption{More visual results on luggage images. The error to the ground truth is computed using the conditional mean $\E[\xbm|\ybm]$, and the variance corresponds to  per-pixel standard deviation.  It is evident that the ill-posed nature of the reconstruction task has a direct impact on the diversity of the generated samples, and the variances are highly correlated with the reconstruction errors.
	}
	\label{Fig:Lug_MV_sup}
\end{figure*}

\medskip\noindent
\textbf{More Results for Uncertainty Quantification.} Fig.~\ref{Fig:ckc_mv1_sup}-\ref{Fig:Lug_MV_sup} shows additional numerical validation that DOLCE is able to quantify uncertainty by estimating the variances directly. Since a well-calibrated model indicates larger variance in areas of larger absolute error, variance can be used as a proxy for reconstruction error in the absence of ground truth.

\section{3D Segmentation Results}

\begin{figure*}[t!]
	\centering
	\includegraphics[width=0.96\textwidth]{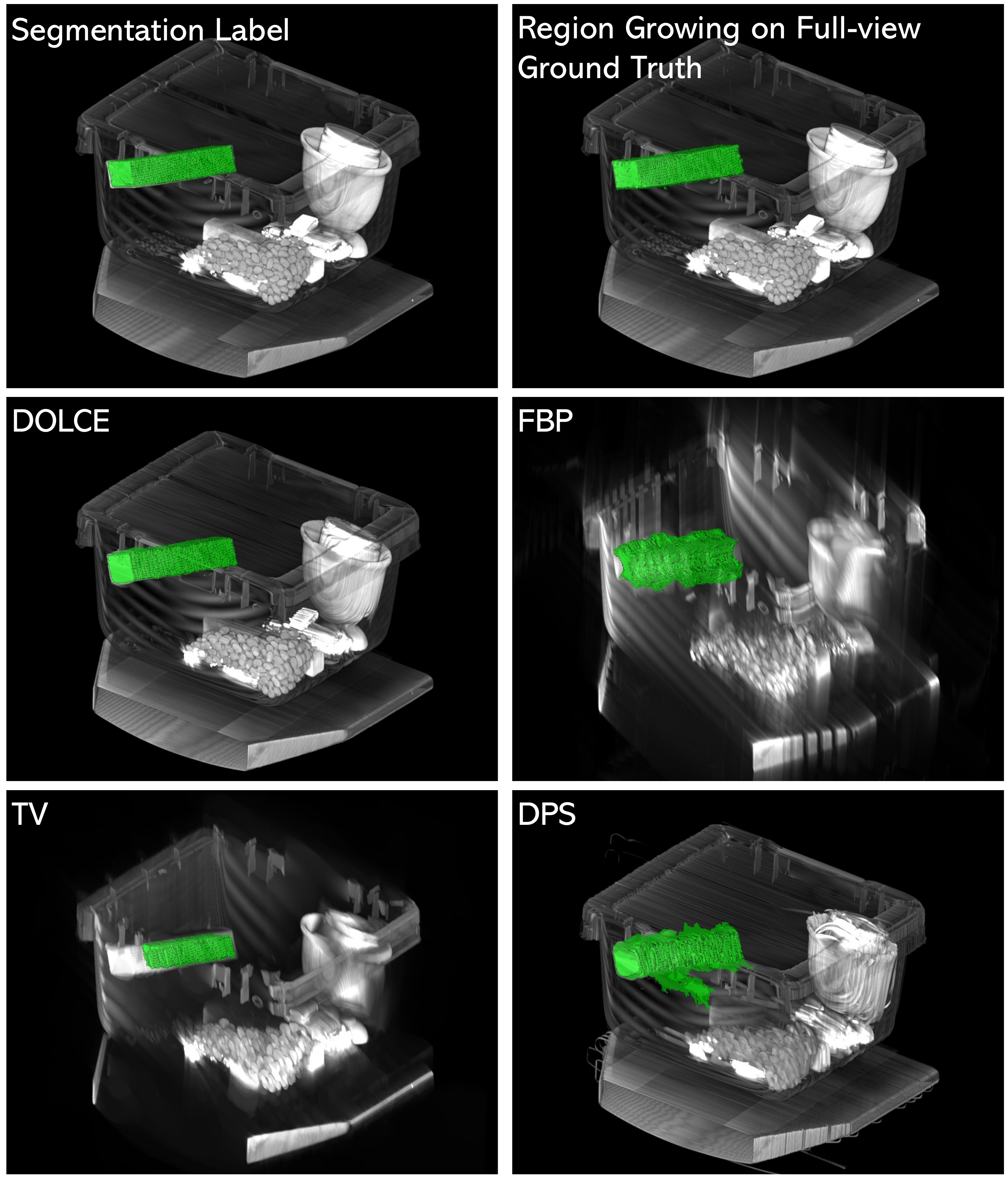}
	\caption{\textbf{More 3D Segmentation Results on Test Bag 2:}  We use a region growing 3D segmentation in all cases and the resulting segmentations are highlighted in color, against a 3D rendering of the 274 reconstructed 2D slices using $\theta_{\text{max}}=$60$^{\circ}$.
	}
	\label{Fig:seg2_sup}
\end{figure*}

\begin{figure*}[t!]
	\centering
	\includegraphics[width=0.96\textwidth]{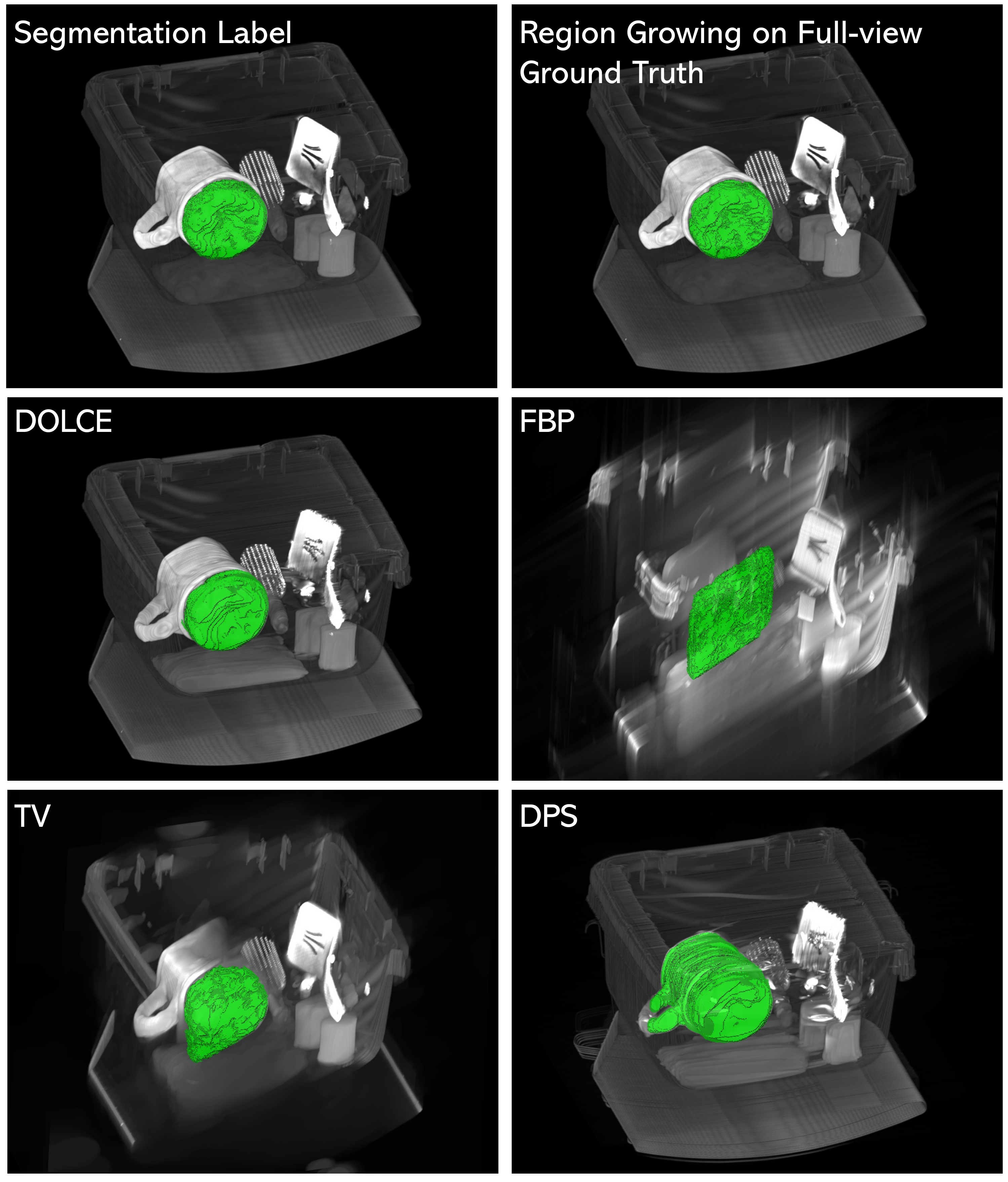}
	\caption{\textbf{More 3D Segmentation Results on Test Bag 3:}  We use a region growing 3D segmentation in all cases and the resulting segmentations are highlighted in color, against a 3D rendering of the 274 reconstructed 2D slices using $\theta_{\text{max}}=$60$^{\circ}$.
	}
	\label{Fig:seg3_sup}
\end{figure*}

\begin{figure*}[t!]
	\centering
	\includegraphics[width=0.96\textwidth]{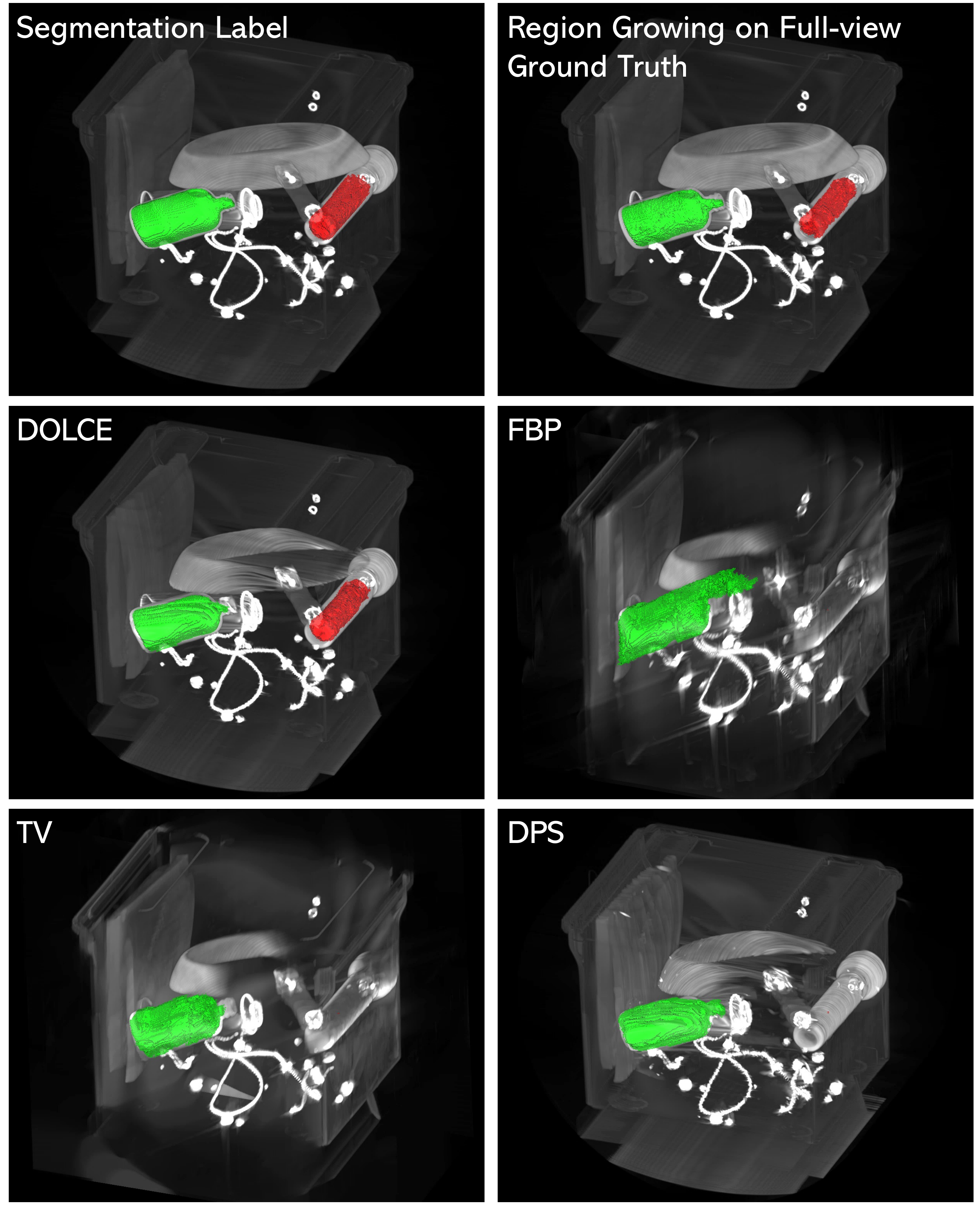}
	\caption{\textbf{More 3D Segmentation Results on Test Bag 4:}  We use a region growing 3D segmentation in all cases and the resulting segmentations are highlighted in color, against a 3D rendering of the 268 reconstructed 2D slices using $\theta_{\text{max}}=$60$^{\circ}$.
	}
	\label{Fig:Fig:seg4_sup}
\end{figure*}

\begin{figure*}[t!]
	\centering
	\includegraphics[width=0.96\textwidth]{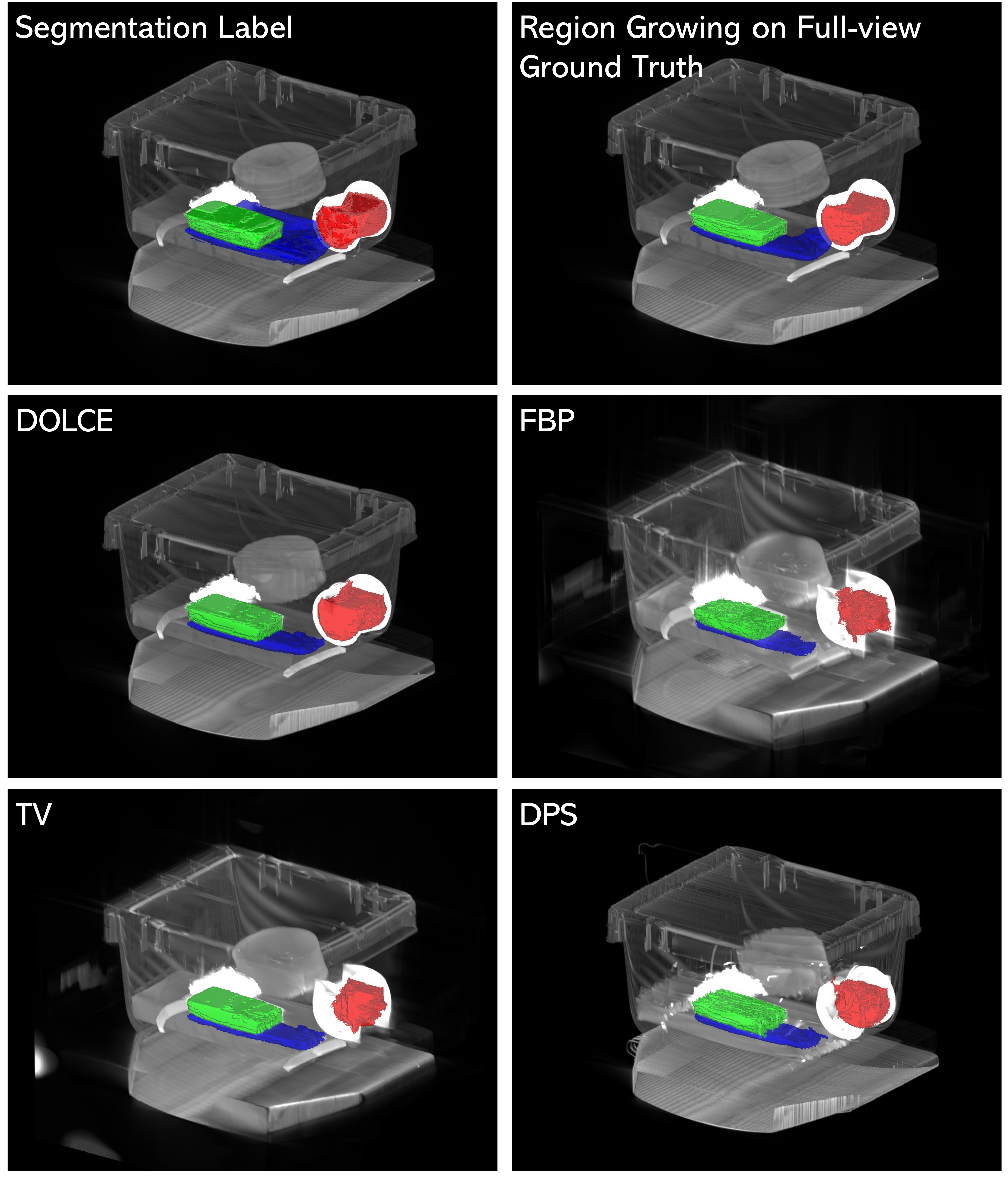}
	\caption{\textbf{More 3D Segmentation Results on Test Bag 5:}  We use a region growing 3D segmentation in all cases and the resulting segmentations are highlighted in color, against a 3D rendering of the 274 reconstructed 2D slices using $\theta_{\text{max}}=$90$^{\circ}$.
	}
	\label{Fig:seg5_sup}
\end{figure*}

\begin{figure*}[t!]
	\centering
	\includegraphics[width=0.96\textwidth]{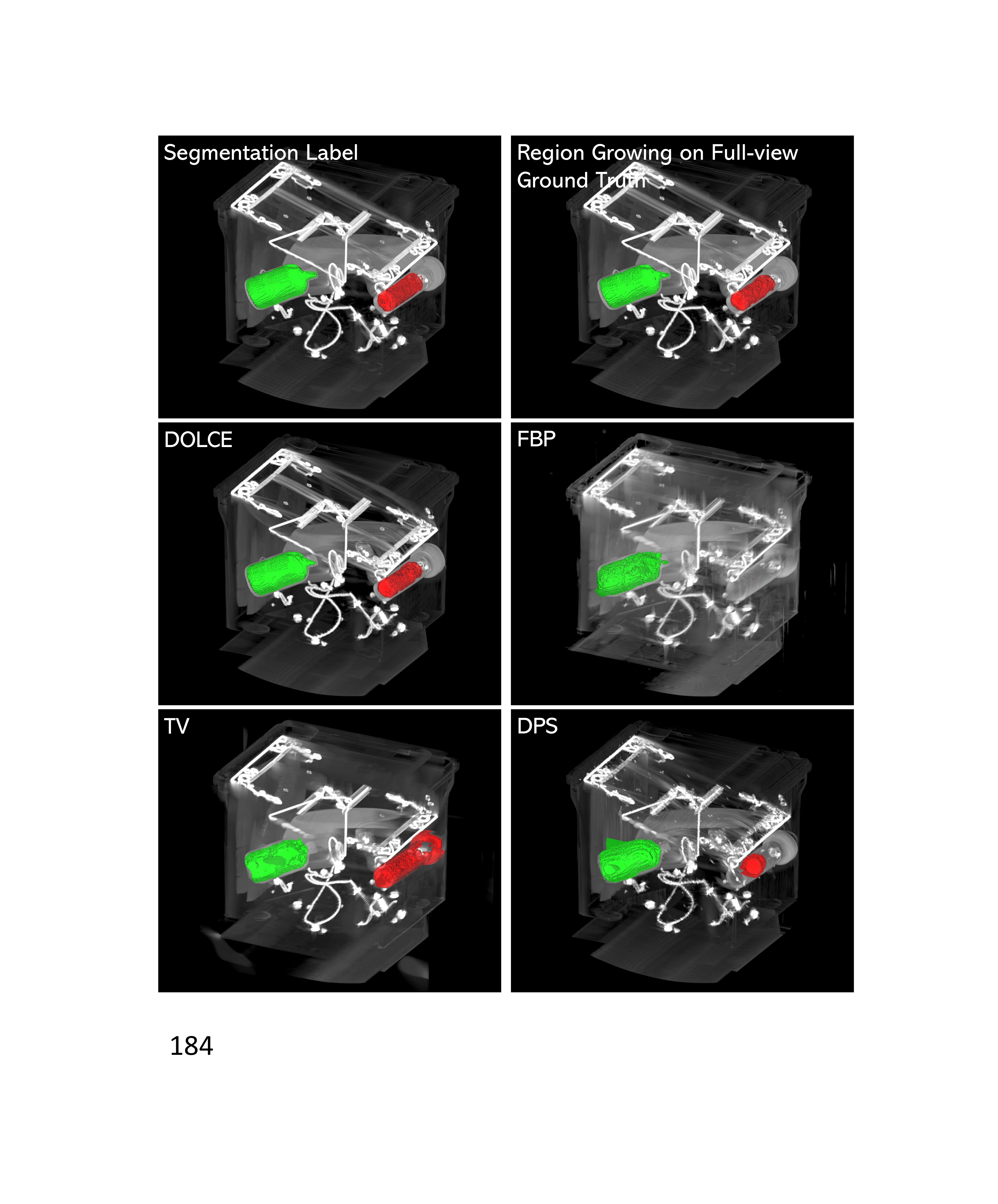}
	\caption{\textbf{More 3D Segmentation Results on Test Bag 6:}  We use a region growing 3D segmentation in all cases and the resulting segmentations are highlighted in color, against a 3D rendering of the 268 reconstructed 2D slices using $\theta_{\text{max}}=$90$^{\circ}$.
	}
 	\label{Fig:seg6_sup}
\end{figure*}

\noindent
We presents additional segmentation results on the reconstruction 2D slices obtained from our DOLCE in Fig~\ref{Fig:seg2_sup}-~\ref{Fig:seg6_sup}. The purpose of these experiments are to evaluate how quality affects object segmentation. In specific, we use a popular region growing
segmentation similar to the method used in~\cite{Kim.etal2015a}, which is a simplified version of the method in~\cite{Wiley.etal2012}, with a randomly chosen starting position and a fixed kernel size. The luggage dataset contains segmentation labels of objects of interest, and the evaluation focuses on how well each segmentation extracts the labeled object. We reconstruct all slices of each bag through the proposed method and combine them into a single bag in 3D. Then we run the region growing in 3D at multiple, hand-tuned parameter settings (intensity threshold ranging from $0.0001$ to $0.02$), and reported the results from the best performing setting. This is done as some reconstruction results are in poor quality and sensitive to the threshold. We compare the segmentations obtained using our method to the segmentation labels as reference, and those obtained using full-view ground truth, FBP, TV and DPS, respectively. It is evident to observe that our proposed reconstruction segments the objects of interest very similar to the ground truth images, than compared to using baseline methods for reconstruction.
\newpage




\end{document}